\documentstyle[12pt,fullpage,epic,eepic]{article}
\begin{document}

\Large
\centerline{\bf Multigrid High Order Mesh Refinement Techniques}

\normalsize
\vspace{1. cm}
\centerline{Thomas L. Beck}
\centerline{\it Department of Chemistry}
\centerline{\it University of Cincinnati}
\centerline{\it Cincinnati, OH 45221-0172}
\centerline{email: becktl@email.uc.edu}
\centerline{\today}

\vspace{1. cm}
\large
\centerline{\bf Abstract}
\vspace{.3 cm}
\normalsize

A method for performing high order mesh refinement multigrid 
computations is presented. The Full Approximation Scheme
(FAS) multigrid technique is utilized for a sequence of
nested patches of increasing resolution. Conservation
forms are generated on coarse scales
by additional defect correction terms
which counter the local excess fluxes at the boundaries.  
Formulas are given for arbitrary order, extending 
the existing technique of Bai and
Brandt.  Test calculations are presented for a singular source
in three dimensions which illustrate the multigrid 
convergence properties, numerical accuracy, and correct order 
of the approach.  Applications for 
all electron quantum chemical 
computations are discussed. 

\newpage
\section{Introduction}

Many if not most computational physics and chemistry problems
require consideration of a large range of length scales. 
In protein folding, both short and long ranged interactions 
and competitions between them
lead to the final configuration of the molecule.\cite{wolynes,dill} 
For the interaction between a protein and a nucleic
acid or a charged interface, specific ionized groups 
may contribute significantly to the binding, while 
distant regions of the molecule have less importance.\cite{honig}  
If one studies
the electronic structure of large molecules, there is a 
concentration of electron density around the atomic
nuclei and between atoms in the chemical bonding regions,
while often large portions of space exhibit very low 
and smoothly varying density.\cite{friesner}  
In numerical simulations
of fluid dynamics, there may exist 
specified regions which 
require a higher resolution treatment locally.\cite{michelsen}

Multiscale methods provide one approach for tackling 
computational problems exhibiting a range of length
scales.\cite{brandt,briggs}
These methods were developed in order to overcome 
convergence problems in iterative solutions to partial
differential equations.  By utilizing approximations
from coarser grids, components of the error on a wide
range of length scales can be decimated, 
typically leading to 
linear scaling computing time with system size. 
The underlying differential equations
can be represented in various ways including finite 
differences and finite elements. For the present method,
high order finite difference representations
are employed. 

Based on the physical examples given above and many others
requiring variable resolution, 
it is appropriate 
to develop mesh refinement strategies in which fine
gridding is focused only in those 
spatial regions which require
it.\cite{bai/brandt}  
The goal is to maintain the linear scaling 
property of the multigrid method while minimizing the
prefactor in the scaling relation. The computational
philosophy adopted here is to generate a sequence of
nested {\it regular} grids; this strategy allows one to use
the existing multigrid routines over the mesh refinement
patches with no significant changes. It also allows 
for the implementation of accurate high order difference
equations on a composite mesh. This approach is to
be contrasted with the many curved grid generation 
techniques\cite{gridgen} which have been 
widely applied in engineering
applications. 

Bai and Brandt\cite{bai/brandt} 
addressed several pivotal issues concerning
extension of the FAS multigrid technique for locally 
refined meshes.  First, they developed a $\lambda$-FMG 
altorithm in order to restore linear scaling behavior 
which can be lost when many levels of refinements are
used and thus the coarser global grids are themselves visited 
in a way which scales linearly with the number of levels. 
The work to accuracy exchange rate $\lambda$ is the Lagrange
multiplier for the grid optimization equations. Second, they 
determined that the same interpolation order can be used
at the mesh boundaries as is used over the rest of the domain.
Third, they found that local relaxation sweeps near structural
singularities (to be differentiated from source singularities)
can restore convergence rates to those observed away from
the singularities.  Finally, a second order conservative differencing
method was developed for interior source singularities, when
it was found that the only important factor for obtaining 
accurate solutions far from the singularity was to correctly
reproduce the source strength around that singularity. These
mesh refinement techniques were tested on several model two
dimensional problems. 

In the present work, the Bai and Brandt 
conservative differencing method is extended to high orders 
and three dimensions. The 
method is then successfully tested on a 
three dimensional
source singularity, the Coulomb potential. 
First, a review of the FAS multigrid method is given. Next, 
a simple approach for generating the high order
formulas for the Laplacian and the interpolation near
the boundaries is reviewed. Then, the conservative 
FAS differencing forms
are derived from general considerations of balancing the 
fluxes at the boundaries. The three dimensional conservative forms
are obtained simply by locally averaging the one dimensional
fluxes over surfaces at the boundaries. With the inclusion of the 
high order flux corrections, the summation of the defect
correction over the mesh refinement is zero to machine 
precision. The boundary corrections lead to accuracy within
the patch which is the same as that for a 
high order uniform fine
mesh covering the whole domain; with no corrections,
serious errors occur over the whole composite
domain. The accuracy outside the 
refinement zone is improved over that for the high order
uniform coarse mesh over the whole domain. 
The correct high order behavior is thus
obtained over the 
entire composite domain. The only exceptions are the points one
grid spacing away from the singularity, where high order
is not obtained on uniform nor composite meshes. 

High orders are often required in order to obtain numerically 
accurate results on three dimensional grids of 
reasonable size. While the methods derived here are
general for elliptic problems, part of our motivation stems 
from the recent development, along with other groups,  
of {\it ab initio} multigrid methods
for quantum chemistry.\cite{gygi,bernholc,beck}  In our 
all electron approach, all particles 
are represented numerically on the grid, including the 
nuclei. 
In previous work, we have 
carried out density functional calculations on atoms
and small molecules on uniform grids, and it is clear 
that the majority of the numerical errors originate from 
the regions around the nuclei. These errors are due both
to the finite size of the nucleus on the grid and to the 
poor numerical representation of the core electron
orbitals. 
The methods developed in this paper will 
be incorporated in 
our electronic structure codes both for the Poisson solver
and for the eigenvalue solver. 

\section{Full Approximation Scheme Algorithm}

The FAS multigrid technique allows for solution of nonlinear
problems and is ideally suited for the mesh refinement methods
presented below.\cite{brandt,bai/brandt,abeval}  
A two dimensional 
schematic of the composite mesh
examined here is presented in Figure 1, with two nested
patches within a full domain. In the test 
calculations here, four or five
levels were employed with the coarsest three covering
the entire domain (see Figure 2).
The Poisson equation to be solved on the finest 
patch of the four level problem is written as:

\begin{equation}
L^{h4} U^{h4} = f^{h4}
\end{equation} 

\noindent
For this case, $L^{h4}$ is the finite difference Laplacian on 
the finest scale, $U^{h4}$ is the exact grid solution
for the potential on that scale, 
and $f^{h4}$ is $-4\pi$ times the charge
density for a three dimensional problem.   The current
approximation to the exact solution will be written in 
lower case: for example the approximate $h4$ solution
is $u^{h4}$. 
On the next 
coarser level $h3$, the level $h4$ patch covers only a
portion of the level $h3$ domain. The 
equation to be solved on the $h3$ level is:

\begin{equation}
L^{h3} U^{h3} = I{_{h4}^{h3}}f^{h4} + \tau^{h3},
\end{equation} 

\noindent
where  
$I{_{h4}^{h3}}$ is the full restriction operator which 
performs a weighted local average of the fine scale 
function, and
$\tau^{h3}$ is the level $h3$ defect correction
given by:

\begin{equation}
\tau^{h3} = L^{h3}I{_{h4}^{h3}}u^{h4} - I{_{h4}^{h3}}L^{h4}u^{h4}.
\end{equation} 

\noindent
The fine scale function can then be corrected
as follows:

\begin{equation}
u^{h4} \leftarrow u^{h4} + I{_{h3}^{h4}}(u^{h3}-
I{_{h4}^{h3}}u^{h4})
\end{equation}

\noindent
and further iterations are subsequently performed on the
$h4$ level. 
Here $I{_{h3}^{h4}}$ is the interpolation operator. 
One way to understand the function of the defect 
correction is to observe that, if the exact solution
$U^{h4}$ were passed to level $h3$, no correction 
would be made. That is, the defect correction 
causes the level $h3$ equation to `optimally mimic'
the level $h4$ problem.  Note that $\tau^{h3}$ is
only defined over the coarse grid points within
the interior region of the 
$h4$ level patch, with zero values outside. 

When using multiple scales, the defect correction 
includes an additional contribution from 
the previous scale. Here 
the example is given of the level $h2$ defect correction
computed during the final level $h4$ V cycle:  

\begin{equation}
\tau^{h2} = L^{h2}I{_{h3}^{h2}}u^{h3} - I{_{h3}^{h2}}L^{h3}u^{h3}
+I{_{h3}^{h2}}\tau^{h3}.
\end{equation} 

\noindent
By performing these coarse grid correction cycles recursively
to coarser and coarser scales, errors of all wavelengths can
be effectively removed with only several iterations necessary
on the fine scale. 

Before deriving the high order conservative forms for 
the composite mesh computations, a
simple general procedure for obtaining the 
required high order Laplacian
and interpolation operators is summarized. 

\section{Direct Method for High Order Formulas}

For completeness, a summary is given here of a general
method for developing any of the formulas
required in a high order multigrid method. 
Hamming\cite{hamming} outlines a direct method for 
obtaining numerical
formulas of arbitrary order. Sample a function at $N$ points.
Define the Lagrange sample
polynomials:

\begin{equation}
\pi{_i}(x) = (x-x_1)(x-x_2) \cdots (x-x_{i-1})
(x-x_{i+1}) \cdots (x-x_{N}), 
\end{equation}

\noindent
which are of order $N-1$ in $x$. Then 

\begin{equation}
\pi{_i}(x_j) = 0,
\end{equation}

\noindent
and

\begin{equation}
\pi{_i}(x_i) \neq 0.
\end{equation}

\noindent
A polynomial which passes through the $N$ sample 
points $y_i$ is: 

\begin{equation}
P_{N-1}(x) = \sum_i^N y_i \left[ \frac{\pi_i(x)}{\pi_i(x_i)} 
\right] . 
\end{equation}

Expand the sample polynomial as follows:

\begin{equation}
\pi_i(x) = c_{1,1} + c_{1,2}x + \cdots + c_{i,N}x^{N-1},
\end{equation}

\noindent
where $c_{i,N} = 1$.  Then it is easy to see that the 
product of the coefficient matrix and the Vandermonde 
matrix of the sample points (second matrix below)
is the matrix $[\pi_i(x_j)]$:

\begin{equation}
{\left[
\begin{array}{cccc}
c_{1,1} & c_{1,2} & \cdots & c_{1,N} \\
c_{2,1} & c_{2,2} & \cdots & c_{2,N} \\
\vdots  & \vdots  & \vdots & \vdots  \\
c_{N,1} & c_{N,2} & \cdots & c_{N,N} 
\end{array} 
\right]}
{\left[
\begin{array}{cccc}
1       &      1  & \cdots &  1      \\
x_1     &     x_2 & \cdots &  x_N    \\
x_1^2   &  x_2^2  & \cdots & x_N^2   \\
x_1^{N-1}&x_2^{N-1}&\cdots & x_N^{N-1} 
\end{array}
\right]}  = [\pi_i(x_j)] . 
\end{equation}

\noindent
Thus, except for a normalization factor of $\pi_i(x_i)$
for each row of the coefficient matrix,  the left hand
matrix is the inverse of the Vandermonde matrix. 

In the direct method, Hamming shows that there is a 
simple connection between the Vandermonde matrix, the
desired weights for the approximation, and a vector
of `moments':  

\begin{equation}
{\bf X w = m}
\end{equation}

\noindent
These moments result from allowing the 
operation of interest  
to act on the sequence of functions:
$ 1, x, x^2, x^3, \ldots x^{N-1}$.   For example, for
the second derivative operator centered at
$x = 0$, the first four elements of the moment vector
are 0, 0, 2, 0.  Similarly, moments can also be obtained for 
the operations of integration and interpolation.  For integration,
the moments are the integrals of $1, x, x^2, \ldots$
over the sampling domain, and 
for interpolation, they are these elementary polynomials 
themselves.  

Since the normalized 
version of the coefficient matrix in Eqn.\ 11 is 
the inverse of $\bf X$, the weights for 
each of the approximations in the multigrid process
can be calculated
to any desired order by one matrix-vector multiply.
The normalized coefficient matrices for $N$ sampling
points are obtained by expanding the sampling
polynomial (Eqn.\ 10) and dividing each row by the 
normalizing factor $\pi_i(x_i)$. 
They are termed `universal matrices' by Hamming
due to their generality, {\it i.e.} they 
depend only on the sampling points, not on the formula
to be approximated. The matrices are 
tabulated in Ref.\ \cite{hamming}
up to seven sampling points, which allows for 
computation up to the 6th order Laplacian.  
In the present work,
simple C codes have been written which 
generate the eight, nine, 
{\it etc.}  point
matrices as well. The weight vectors for the
Laplacians through 8th order are given in Table 1. 
The three dimensional
versions are generated from the sum of the three orthogonal 
$x, y, z$ axes.
We have utilized Laplacians up to 8th order
in previous multigrid work on uniform domains.\cite{beck}
The interpolation weight vectors for even numbers of
sampling points are listed in Table 2
through 8th order. 
Similary, formulas for any higher orders can be obtained from 
the universal matrices. 
The high order
interpolation formulas only need be used when setting the 
function values on and outside the refinement boundaries
which are fixed during iterations over the patch.  It is
essential that the order of the interpolation match 
the order of the Laplacian at these boundaries. 
Lower
order interpolations are adequate during the rest of 
the multigrid processing. 

\section{High Order Conservative Forms}

When solving for the potential on coarser scales which contain
a mesh refinement patch at the next finer level, 
it is clear from Eqn.\ 2 that, if 
the sum of $\tau$ over the interior domain is not zero, 
additional sources have been introduced. This is in fact 
the case, which can be shown by examination of the $\tau$
terms in a one dimensional example; most of the interior
terms do cancel, but nonzero contributions remain at the
patch boundaries.  The terms which remain are of the form
of one dimensional flux operators. Without correcting for
these new sources, the solution will be polluted over 
the whole domain. The method of Bai and Brandt corrects
for these sources by introducing local opposing fluxes at
the boundary. In this section, their second order
method is extended to high orders. 

First, the problem is illustrated schematically by 
using a continuous notation (in the grid notation, all
integrals go over to sums).  The coarse
scale is labelled by $H$ and the fine by $h$.  
It is desired to satisfy:

\begin{equation}
\int_D \tau^H dV = 0.
\end{equation}

\noindent
where the integration is over the whole patch domain $D$, 
including the boundaries.  However, it is true that

\begin{equation}
\int_I \tau_{int}^H dV \neq 0.
\end{equation}

\noindent
This integration is only over the interior region 
of the patch.  Therefore:

\begin{equation}
\int_D \tau^H dV = 
\int_I \tau_{int}^H dV +
\int_S \tau_{b}^H dV  = 0.
\end{equation}

\noindent
Here $\tau_b^H$ is a boundary term designed to oppose locally
the additional terms due to nonconservation of source
and the $S$ integration is over a narrow strip at the surface. 
This implies:

\begin{equation}
 \int_S \tau_{b}^H dV  =
- \int_I \tau_{int}^H dV .
\end{equation}

\noindent
The form of $\tau_{int}^H$ is the difference of the Laplacian
acting on the coarse scale function minus a local average
of the Laplacian acting on the fine scale function, Eqn.\ 3.
Therefore, converting a volume integral into a surface
integral:

\begin{equation}
 \int_S \tau_{b}^H dV =
- \int_I [(\nabla^2)^H u^H - \langle (\nabla^2)^h u^h \rangle] dV 
= - \int_{\Omega} [\nabla_b^H u^H - 
\nabla_b^h u^h] d{\bf \sigma} .
\end{equation}

\noindent
The brackets $\langle\rangle$ signify a 
local average (restriction) of the fine scale 
Laplacian acting on the function, 
and the gradient operators $\nabla_b$ are
obtained by noncancellation of terms near the
boundary of the volume
integral. 
The final expression shows that the boundary $\tau_b^H$
generates a flux which locally opposes the flux 
from the additional sources
in the interior. Therefore, after collecting the correct 
units from the two scales, it is apparent that the 
form for $\tau_b^H$ is:

\begin{equation} 
- \tau_{b}^H  H^2 a =   
[\nabla_b^H u^H -  \nabla_b^h u^h] ,  
\end{equation} 

\noindent 
where $H$ is the coarse grid spacing, $a$ is the numerical
prefactor to the Laplacian (see Table 3),  and now 
the gradients are one dimensional operators directed 
outward from the surface (determined below). 
Here the gradients simply represent 
the unitless coefficients since $H^2$ and $a$ have been 
moved to the other side. For example, on a one 
dimensional domain (corresponding to
the 2nd order Laplacian):

\begin{equation}    
[\nabla^H u^H]_i = u_i^H - u_{i+1}^H
\end{equation} 

\noindent
on the left boundary, and 

\begin{equation}    
[\nabla^H u^H]_j = u_{j}^H - u_{j-1}^H
\end{equation} 

\noindent
on the right. 

Since the process of full restriction 
is a weighted 
local average over the fine scale function (27 points
in three dimensions), 
the averaging in 
Eqn.\ 17 can be viewed as follows. First,
average over the direction normal to the boundary surface,
compute the two gradient terms on the rhs of Eqn.\ 18, 
and then average over the other two directions. The
full restriction weights for this process along one
fine scale
dimension are ${\bf w}$ = [1/4 1/2 1/4].  There is no 
requirement for high order restriction operators, as long
as the same restriction method is used consistently. 
Therefore, the coefficients for the 
two gradient terms in Eqn.\ 18 can be determined 
by solving the one dimensional 
problem. 
In two dimensions, the fine scale gradient operator
$\nabla_b^h$ is averaged over three points along the 
boundary line with weights [1/4 1/2 1/4]. 
This yields
a local average of the fine scale flux through the 
boundary. A similar procedure
was followed in the work of Bai and Brandt.\cite{bai/brandt} 
In three 
dimensions, the local flux average is over a square centered on the
location of the coarse scale gradient. The weights 
are 1/4 for the center, 1/8 for the edges, and 1/16 for
the corners. 

The one dimensional version of the flux difference in 
Eqn.\ 18 was solved for high orders by examination of 
the cancellation of terms near the boundary.  The result
for the left hand side of the 
coarse scale gradient on a left boundary is 
given by: 

\begin{equation}
d_{{-n_L}+i} = \sum_{j=0}^{i} c_{{-n_L}+j}\hspace{1cm}  i = 0, n_L - 1,
\end{equation}

\noindent
where $n_L$
is the number of points in the Laplacian to the left of 
the center and the $c_{{-n_L}+j}$ are the Laplacian 
coefficients from Table 1. The rhs side of the gradient 
is antisymmetric with respect to these coefficients. 
For a right side boundary, all the signs are reversed. 

The locally averaged (in one dimension) 
fine scale gradient coefficients are:

\begin{equation}
e_{{-n_L}+i} = \sum_{j=0}^{i-1} 2c_{{-n_L}+j} +  
c_{{-n_L}+i} \hspace{1cm}   i = 0, n_L - 1.
\end{equation}

\noindent
For the fine scale coefficients, the central term always
cancels completely, so both gradient operators
$\nabla_b^H$ and $\nabla_b^h$ are centered about the 
fine grid location one point inside the patch boundary.
All of the coefficients up through 8th order are 
listed in Table 3, and the terms for
a 6th order left boundary are shown in Fig.\ 3 to 
illustrate the locations.
Similarly, conservative forms can be derived for higher
orders if desired. 

\section{Computational Details and Numerical Results}

The computational test case presented here is for the 4th order
form. The three coarsest scales covered the whole domain, while
the finest one or two were nested patches. 
On the three coarsest (full domain) scales, 
the boundaries were set by fixing
the potential at the analytical value for a singular
source in three dimensions $\phi({\bf r})=1/r$.  The boundary was
fixed with one additional term outside the physical 
boundary since the
Laplacian has two terms beyond the center in one dimension. 
Iterations were performed over all the interior 
points of the full 
domain or patch. 
The FAS-MG technique was used in the form of the series of
nested V cycles as shown in Fig.\ 2.  SOR iterations
were employed for all relaxation steps, 
with $\omega = 1.2$. The optimal relaxation parameter was determined
empirically for the high order case. Full weighting restriction
and linear interpolation were used,
except 4th order interpolation was performed
over the patch regions, including the required points
beyond the boundaries.  These points
were set such that the Laplacian 
and defect correction were
defined over the entire interior of the patch.  The boundary 
potential terms for the patches were reset during the correction
step after each visit to 
coarser scales. 

The code was written in C with double precision
arithmetic, utilizing the prescription 
of Ref.\ \cite{numrec} for
dynamic memory allocation.  The test calculations
were run on a Pentium 133 MHz processor laptop with 40Mb of 
RAM, requiring a total of roughly 15 relaxation sweeps on the fine
scale and 3 seconds of total processor time for 
convergence. The `exact'
grid results were obtained by repeated loops around the 
final V cycle of the FAS procedure, until the residuals 
were on the order of machine precision zero.  
The coarsest (full domain) scale had 5 points on a side, 
the next two finer scales 9 and 17,
and the two nested patches both had 9 points on a side.  
To examine the order of the method, computations
were performed on a full domain coarse grid corresponding
to the that of the composite mesh (level $h3$), 
followed by one finer full domain grid with the 
spacing halved. The accuracy of the composite mesh method
was then determined by comparison with the high order coarse
and fine uniform grid calculations used in the determination
of the order. 

The increased accuracy obtained using fourth order equations
{\it vs.} second order is displayed for uniform domain
computations in Figure 4, in 
which the absolute errors of the solution are presented
away from the singularity. That the 
fourth order Laplacian leads to fourth order behavior was
confirmed in the uniform grid computations described above. 
Except for the set of points one grid spacing away from
the singularity, the correct order is obtained over the 
entire domain. 

Then computations were performed on the four level composite
domain with a single refinement patch centered at the origin. 
To test the effect of the boundary correction on conservation,
the integral of the defect correction over the refinement patch 
was computed with 
and without the boundary terms. Without the boundary correction,
the integral was 0.8 in magnitude, while 
with the boundary terms, the integral was zero to double
precision accuracy. The impact of the conservative boundary
correction on the accuracy of the solution is apparent in
Figure 5; serious errors are incurred over the whole domain
in the absence of the boundary corrections. 
The accuracy of the method can be determined by comparison
with the separate fine and coarse uniform domain results. 
Figure 6 shows that the accuracy within the patch is 
virtually identical to that for the uniform fine domain
results. In Figure 7, the errors outside the refinement 
patch are displayed.  The refinement mesh leads to 
increased accuracy outside the refinement on the coarser
level in comparison with the uniform coarse level results. 
The numerical results are presented in Table 4. Finally,
test computations were also successfully performed on a five level
problem with two nested refinement patches. The 
resulting potential is plotted in Figure 8 to illustrate
the accuracy of the method in relation to the numerical errors. 
These numerical results thus confirm that the conservative mesh
refinement technique developed here leads to results 
of the correct high order within the refinement region,
while increasing the accuracy on the coarse
domain outside the refinement zone. 

\section{Summary}

A general technique has been presented for carrying out high
order mesh refinement multigrid calculations. The 
FAS method for composite domains was first summarized.
Then, Hamming's direct method for generating high order
formulas was outlined.  Both the high order Laplacian 
and interpolation coefficients were obtained from the 
universal matrices of the direct method.  Since the 
sum of the defect correction over the interior of the
patch is nonzero, high order conservative forms were
derived by analysis of the one dimensional problem.  
The two and three dimensional forms can be obtained
by averaging locally over three points on 
a line or nine points on a square, respectively. The 
new method was successfully
tested for the fourth order case on a Poisson problem,
the source singularity in three dimensions.  

The high order mesh refinement methods should allow for 
accurate computations on three dimensional domains which
require a range of length scales. We are developing a 
quantum chemical Density Functional Theory (DFT) 
multigrid method
for {\it ab initio} calculations.\cite{beck} 
So far, our 
fully numerical three dimensional calculations
have been performed on uniform domains, treating both the electrons
and nuclei with the high order 
grid approximations. As a test computation, 
we examined the CO molecule, and obtained good results in 
all electron computations. However, it is apparent from those
results that the crude treatment of the nuclei and the core
electrons limits the accuracy of the method. We plan to 
incorporate the high order composite mesh techniques into
the quantum chemistry method to obtain more accurate results
in the region of the nuclei, and to investigate the impact
of those improvements on the eigenfunctions, eigenvalues, and
total molecular energies.  Beyond the quantum chemical 
applications, these methods should prove helpful in other large
scale electrostatics calculations in biophysics and in 
multigrid fluid dynamics computations, especially for cases 
where increased resolution is required only over small local
regions of space. 

\vskip 0.5in
\Large
\noindent
{\bf Acknowledgments}
\normalsize

\noindent
I thank Achi Brandt, Dov Bai, and Michael Merrick for
many helpful discussions. 
I would like to acknowledge the support of NSF grant 
CHE-9632309. I also thank Daan Frenkel and Bela Mulder
for support during a sabbatical leave at the FOM 
Institute in Amsterdam during the fall of 1996.



\newpage


\begin{table}
\centerline
{\begin{tabular}{|r|r|r|r|r|r|r|r|} \hline
Points & Order & Prefactor & 
\multicolumn{5}{c|}{Coefficients} \\ \hline
N=3 & 2nd & 1 & & & & 1 & -2  \\ \hline
N=5 & 4th & 12 & & & -1 & 16 & -30  \\ \hline
N=7 & 6th & 180 & & 2 & -27 & 270 & -490  \\ \hline
N=9 & 8th & 5040 &-9 &128 & -1008 & 8064 & -14350  \\ \hline
\end{tabular} }
\caption{Coefficients for the Laplacian.  One side plus 
the central point are shown. 
Each coefficient term should be divided by the prefactor.
The Laplacian is symmetric about the central point.}
\end{table}

\begin{table}
\centerline
{\begin{tabular}{|r|r|r|r|r|r|r|} \hline
Points & Order &   Prefactor &
\multicolumn{4}{c|}{Coefficients} \\ \hline
N=2 & 2nd & 2 & & &  & 1  \\ \hline
N=4 & 4th & 16 & & & -1 & 9   \\ \hline
N=6 & 6th & 256 & & 3 & -25 & 150   \\ \hline
N=8 & 8th & 2048 &-5 & 49 & -245 & 1225 \\ \hline
\end{tabular} }
\caption{Coefficients for interpolation.  One side 
of the symmetric weight vector is 
shown. 
Each coefficient term should be divided by the prefactor.}
\end{table}

\begin{table}
\centerline
{\begin{tabular}{|r|r|r|r|r|r|r|r|} \hline
Level &Points & Order &   Prefactor &
\multicolumn{4}{c|}{Coefficients} \\ \hline
H & N=2 & 2nd & 1 & & &  & 1  \\ 
h &  &  &  & & &  & 1  \\ \hline
H& N=4 & 4th & 12 & & & -1 & 15   \\ 
h&  &  &  & & & -1 & 14   \\ \hline
H& N=6 & 6th & 180 & & 2 & -25 & 245   \\ 
h&  &  &  & & 2 & -23 & 220   \\ \hline
H& N=8 & 8th & 5040 &-9 & 119 & -889 & 7175 \\ 
h&  &  &  &-9 & 110 & -770 & 6286 \\ \hline
\end{tabular} }
\caption{Coefficients for conservative forms.  
One side is 
shown. Each term on the other half of the gradient
operator has the opposite sign. 
The set of coefficients is for a left 
boundary. All the signs are reversed for a right 
boundary. The locations of the terms 
for a 6th order example are shown 
in Fig.\ 3.}
\end{table}

\begin{table}
\centerline
{\begin{tabular}{|r|r|r|r|r|r|r|} \hline
r & Exact & Grid Exact & Trunc.\ Err.\ & Fine Trunc.\ Err.&
 Coarse Trunc.\ Err.& MG Err. \\ \hline
2 & 0.5 & 0.520842737 & 0.020842737 & 0.020888618 &  & 0.020875723 \\ \hline
4 & 0.25 & 0.257687505 & 0.007687505 & 0.007710007  & 0.010443407 & 0.007726450 \\ \hline
6 & 0.1$\bar{6}$ & 0.167993560 & 0.001326893 & 0.001183652 &  & 0.001389060\\ \hline
8 & 0.125 & 0.126091267 & 0.001091267 & 0.000232278 & 0.003854094 & 0.001232428 \\ \hline
12 & 0.08$\bar{3}$ & 0.0834252199 & 0.0000918866 & 0.0000295553 & 0.0005908833 & 0.0001265184 \\ \hline
16 & 0.0625 & 0.0625022779 & 0.0000022779 & 0.0000072359 & 0.0001151041 & 0.0000094832 \\ \hline
20 & 0.05 & 0.0499998476 & -0.0000001524 & 0.0000023526 & 0.0000347405 & 0.0000016978 \\ \hline
24 & 0.041$\bar{6}$ & 0.0416667925 & 0.0000001258 & 0.0000008825 & 0.0000131217 & 0.0000007387 \\  \hline
\end{tabular} }
\caption{Numerical results for the FAS-MG composite mesh 
computations at
several distances from the singular source. The edge of the 
patch is at $r=8$. The last column is 
for a single V cycle MG computation with a total of 16 relaxation
sweeps on the fine scale. Notice that for the points in the outer
regions of the domain, the single cycle MG errors are not strictly
less than the anomalously small truncation errors for the composite
domain; however, the errors are considerably smaller than the 
truncation errors on the uniform coarse scale domain (they are
of the same magnitude as the uniform 
fine grid results at those points). } 
\end{table}


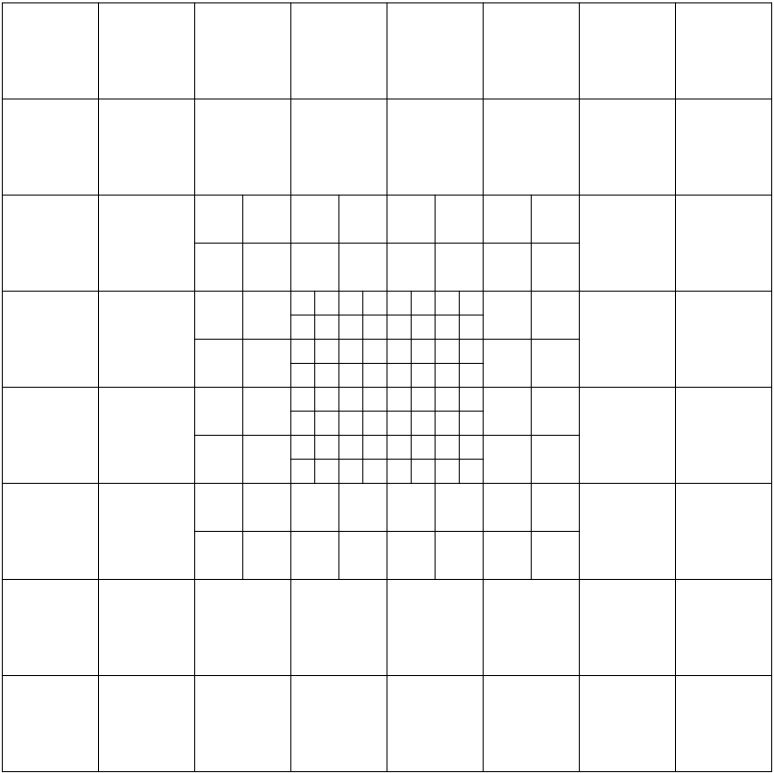
\begin{figure}
\begin{center}
\setlength{\unitlength}{0.00083333in}
\begingroup\makeatletter\ifx\SetFigFont\undefined%
\gdef\SetFigFont#1#2#3#4#5{%
  \reset@font\fontsize{#1}{#2pt}%
  \fontfamily{#3}\fontseries{#4}\fontshape{#5}%
  \selectfont}%
\fi\endgroup%
{\renewcommand{\dashlinestretch}{30}
\begin{picture}(4824,5439)(0,-10)
\drawline(1212,12)(1212,12)
\path(12,5412)(12,612)
\path(612,5412)(612,612)
\path(1212,5412)(1212,612)
\path(1812,5412)(1812,612)
\path(2412,5412)(2412,612)
\path(3012,5412)(3012,612)
\path(3612,5412)(3612,612)
\path(4212,5412)(4212,612)
\path(4812,5412)(4812,612)
\path(12,5412)(4812,5412)
\path(12,4812)(4812,4812)
\path(12,4212)(4812,4212)
\path(12,3612)(4812,3612)
\path(12,3012)(4812,3012)
\path(12,2412)(4812,2412)
\path(12,1812)(4812,1812)
\path(12,1212)(4812,1212)
\path(12,612)(4812,612)
\path(1512,4212)(1512,1812)
\path(2112,4212)(2112,1812)
\path(2712,4212)(2712,1812)
\path(3312,4212)(3312,1812)
\path(1212,3912)(3612,3912)
\path(1212,3312)(3612,3312)
\path(1212,2712)(3612,2712)
\path(1212,2112)(3612,2112)
\path(1962,3612)(1962,2412)
\path(2262,3612)(2262,2412)
\path(2562,3612)(2562,2412)
\path(2862,3612)(2862,2412)
\path(1812,3462)(3012,3462)
\path(1812,3162)(3012,3162)
\path(1812,2862)(3012,2862)
\path(1812,2562)(3012,2562)
\end{picture}
}
\end{center}
\caption{Schematic two dimensional cut through the 
three dimensional composite mesh.} 
\end{figure}

\begin{figure}
\begin{center}
\setlength{\unitlength}{0.00083333in}
\begingroup\makeatletter\ifx\SetFigFont\undefined%
\gdef\SetFigFont#1#2#3#4#5{%
  \reset@font\fontsize{#1}{#2pt}%
  \fontfamily{#3}\fontseries{#4}\fontshape{#5}%
  \selectfont}%
\fi\endgroup%
{\renewcommand{\dashlinestretch}{30}
\begin{picture}(5027,1065)(0,-10)
\path(12,30)(312,330)(612,30)
	(1212,630)(1812,30)(2712,930)
	(3612,30)(4512,930)
\put(4812,30){\makebox(0,0)[lb]{\smash{{{\SetFigFont{14}{16.8}{\rmdefault}{\mddefault}{\updefault}h1}}}}}
\put(4812,330){\makebox(0,0)[lb]{\smash{{{\SetFigFont{14}{16.8}{\rmdefault}{\mddefault}{\updefault}h2}}}}}
\put(4812,630){\makebox(0,0)[lb]{\smash{{{\SetFigFont{14}{16.8}{\rmdefault}{\mddefault}{\updefault}h3}}}}}
\put(4812,930){\makebox(0,0)[lb]{\smash{{{\SetFigFont{14}{16.8}{\rmdefault}{\mddefault}{\updefault}h4}}}}}
\end{picture}
}
\end{center}
\caption{Four level FAS cycle.} 
\end{figure}
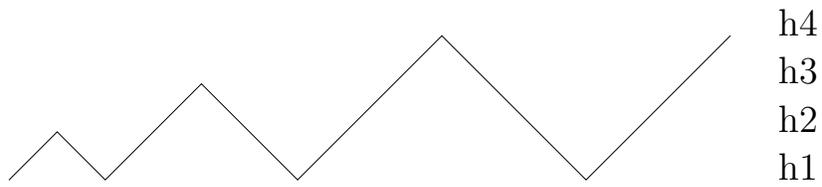

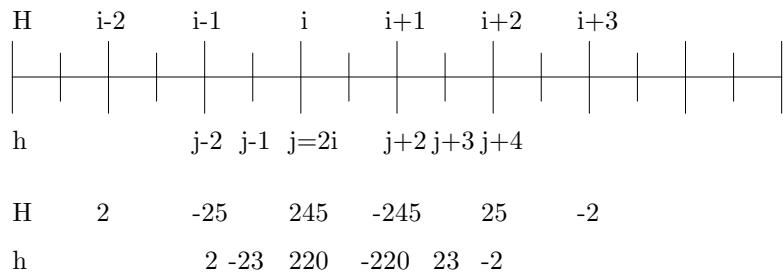
\begin{figure}
\begin{center}
\setlength{\unitlength}{0.00083333in}
\begingroup\makeatletter\ifx\SetFigFont\undefined%
\gdef\SetFigFont#1#2#3#4#5{%
  \reset@font\fontsize{#1}{#2pt}%
  \fontfamily{#3}\fontseries{#4}\fontshape{#5}%
  \selectfont}%
\fi\endgroup%
{\renewcommand{\dashlinestretch}{30}
\begin{picture}(4824,1620)(0,-10)
\path(12,1221)(4812,1221)
\path(12,1446)(12,996)
\path(612,1446)(612,996)
\path(1212,1446)(1212,996)
\path(1812,1446)(1812,996)
\path(2412,1446)(2412,996)
\path(3012,1446)(3012,996)
\path(3612,1446)(3612,996)
\path(4212,1446)(4212,996)
\path(4812,1446)(4812,996)
\path(312,1371)(312,1071)
\path(912,1371)(912,1071)
\path(1512,1371)(1512,1071)
\path(2112,1371)(2112,1071)
\path(2712,1371)(2712,1071)
\path(3312,1371)(3312,1071)
\path(3912,1371)(3912,1071)
\path(4512,1371)(4512,1071)
\put(12,1521){\makebox(0,0)[lb]{\smash{{{\SetFigFont{10}{12.0}{\rmdefault}{\mddefault}{\updefault}H}}}}}
\put(1812,1521){\makebox(0,0)[lb]{\smash{{{\SetFigFont{10}{12.0}{\rmdefault}{\mddefault}{\updefault}i}}}}}
\put(1137,1521){\makebox(0,0)[lb]{\smash{{{\SetFigFont{10}{12.0}{\rmdefault}{\mddefault}{\updefault}i-1}}}}}
\put(537,1521){\makebox(0,0)[lb]{\smash{{{\SetFigFont{10}{12.0}{\rmdefault}{\mddefault}{\updefault}i-2}}}}}
\put(2337,1521){\makebox(0,0)[lb]{\smash{{{\SetFigFont{10}{12.0}{\rmdefault}{\mddefault}{\updefault}i+1}}}}}
\put(3537,1521){\makebox(0,0)[lb]{\smash{{{\SetFigFont{10}{12.0}{\rmdefault}{\mddefault}{\updefault}i+3}}}}}
\put(12,771){\makebox(0,0)[lb]{\smash{{{\SetFigFont{10}{12.0}{\rmdefault}{\mddefault}{\updefault}h}}}}}
\put(1737,771){\makebox(0,0)[lb]{\smash{{{\SetFigFont{10}{12.0}{\rmdefault}{\mddefault}{\updefault}j=2i}}}}}
\put(1437,771){\makebox(0,0)[lb]{\smash{{{\SetFigFont{10}{12.0}{\rmdefault}{\mddefault}{\updefault}j-1}}}}}
\put(1137,771){\makebox(0,0)[lb]{\smash{{{\SetFigFont{10}{12.0}{\rmdefault}{\mddefault}{\updefault}j-2}}}}}
\put(2337,771){\makebox(0,0)[lb]{\smash{{{\SetFigFont{10}{12.0}{\rmdefault}{\mddefault}{\updefault}j+2}}}}}
\put(2637,771){\makebox(0,0)[lb]{\smash{{{\SetFigFont{10}{12.0}{\rmdefault}{\mddefault}{\updefault}j+3}}}}}
\put(2937,771){\makebox(0,0)[lb]{\smash{{{\SetFigFont{10}{12.0}{\rmdefault}{\mddefault}{\updefault}j+4}}}}}
\put(12,321){\makebox(0,0)[lb]{\smash{{{\SetFigFont{10}{12.0}{\rmdefault}{\mddefault}{\updefault}H}}}}}
\put(537,321){\makebox(0,0)[lb]{\smash{{{\SetFigFont{10}{12.0}{\rmdefault}{\mddefault}{\updefault}2}}}}}
\put(1137,321){\makebox(0,0)[lb]{\smash{{{\SetFigFont{10}{12.0}{\rmdefault}{\mddefault}{\updefault}-25}}}}}
\put(1737,321){\makebox(0,0)[lb]{\smash{{{\SetFigFont{10}{12.0}{\rmdefault}{\mddefault}{\updefault}245}}}}}
\put(2262,321){\makebox(0,0)[lb]{\smash{{{\SetFigFont{10}{12.0}{\rmdefault}{\mddefault}{\updefault}-245}}}}}
\put(2937,321){\makebox(0,0)[lb]{\smash{{{\SetFigFont{10}{12.0}{\rmdefault}{\mddefault}{\updefault}25}}}}}
\put(3537,321){\makebox(0,0)[lb]{\smash{{{\SetFigFont{10}{12.0}{\rmdefault}{\mddefault}{\updefault}-2}}}}}
\put(12,21){\makebox(0,0)[lb]{\smash{{{\SetFigFont{10}{12.0}{\rmdefault}{\mddefault}{\updefault}h}}}}}
\put(1212,21){\makebox(0,0)[lb]{\smash{{{\SetFigFont{10}{12.0}{\rmdefault}{\mddefault}{\updefault}2}}}}}
\put(1737,21){\makebox(0,0)[lb]{\smash{{{\SetFigFont{10}{12.0}{\rmdefault}{\mddefault}{\updefault}220}}}}}
\put(2637,21){\makebox(0,0)[lb]{\smash{{{\SetFigFont{10}{12.0}{\rmdefault}{\mddefault}{\updefault}23}}}}}
\put(1362,21){\makebox(0,0)[lb]{\smash{{{\SetFigFont{10}{12.0}{\rmdefault}{\mddefault}{\updefault}-23}}}}}
\put(2187,21){\makebox(0,0)[lb]{\smash{{{\SetFigFont{10}{12.0}{\rmdefault}{\mddefault}{\updefault}-220}}}}}
\put(2937,21){\makebox(0,0)[lb]{\smash{{{\SetFigFont{10}{12.0}{\rmdefault}{\mddefault}{\updefault}-2}}}}}
\put(2937,1521){\makebox(0,0)[lb]{\smash{{{\SetFigFont{10}{12.0}{\rmdefault}{\mddefault}{\updefault}i+2}}}}}
\end{picture}
}
\end{center}
\caption{Locations and values for the coefficients used to 
generate the 6th order conservative form on a 
left boundary. The boundary is located on the coarse scale
$H$ at the point $i$ and on the fine scale $h$ at the 
point $j=2i$. } 
\end{figure}

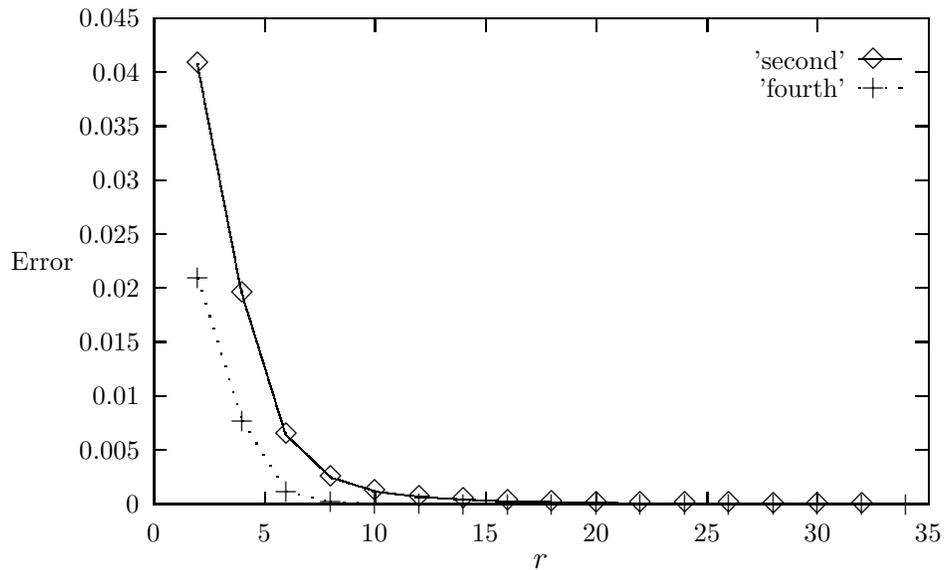
\begin{figure}
\begin{center}
\setlength{\unitlength}{0.240900pt}
\ifx\plotpoint\undefined\newsavebox{\plotpoint}\fi
\sbox{\plotpoint}{\rule[-0.200pt]{0.400pt}{0.400pt}}%
\begin{picture}(1500,900)(0,0)
\font\gnuplot=cmr10 at 10pt
\gnuplot
\sbox{\plotpoint}{\rule[-0.200pt]{0.400pt}{0.400pt}}%
\put(220.0,113.0){\rule[-0.200pt]{292.934pt}{0.400pt}}
\put(220.0,113.0){\rule[-0.200pt]{0.400pt}{184.048pt}}
\put(220.0,113.0){\rule[-0.200pt]{4.818pt}{0.400pt}}
\put(198,113){\makebox(0,0)[r]{0}}
\put(1416.0,113.0){\rule[-0.200pt]{4.818pt}{0.400pt}}
\put(220.0,198.0){\rule[-0.200pt]{4.818pt}{0.400pt}}
\put(198,198){\makebox(0,0)[r]{0.005}}
\put(1416.0,198.0){\rule[-0.200pt]{4.818pt}{0.400pt}}
\put(220.0,283.0){\rule[-0.200pt]{4.818pt}{0.400pt}}
\put(198,283){\makebox(0,0)[r]{0.01}}
\put(1416.0,283.0){\rule[-0.200pt]{4.818pt}{0.400pt}}
\put(220.0,368.0){\rule[-0.200pt]{4.818pt}{0.400pt}}
\put(198,368){\makebox(0,0)[r]{0.015}}
\put(1416.0,368.0){\rule[-0.200pt]{4.818pt}{0.400pt}}
\put(220.0,453.0){\rule[-0.200pt]{4.818pt}{0.400pt}}
\put(198,453){\makebox(0,0)[r]{0.02}}
\put(1416.0,453.0){\rule[-0.200pt]{4.818pt}{0.400pt}}
\put(220.0,537.0){\rule[-0.200pt]{4.818pt}{0.400pt}}
\put(198,537){\makebox(0,0)[r]{0.025}}
\put(1416.0,537.0){\rule[-0.200pt]{4.818pt}{0.400pt}}
\put(220.0,622.0){\rule[-0.200pt]{4.818pt}{0.400pt}}
\put(198,622){\makebox(0,0)[r]{0.03}}
\put(1416.0,622.0){\rule[-0.200pt]{4.818pt}{0.400pt}}
\put(220.0,707.0){\rule[-0.200pt]{4.818pt}{0.400pt}}
\put(198,707){\makebox(0,0)[r]{0.035}}
\put(1416.0,707.0){\rule[-0.200pt]{4.818pt}{0.400pt}}
\put(220.0,792.0){\rule[-0.200pt]{4.818pt}{0.400pt}}
\put(198,792){\makebox(0,0)[r]{0.04}}
\put(1416.0,792.0){\rule[-0.200pt]{4.818pt}{0.400pt}}
\put(220.0,877.0){\rule[-0.200pt]{4.818pt}{0.400pt}}
\put(198,877){\makebox(0,0)[r]{0.045}}
\put(1416.0,877.0){\rule[-0.200pt]{4.818pt}{0.400pt}}
\put(220.0,113.0){\rule[-0.200pt]{0.400pt}{4.818pt}}
\put(220,68){\makebox(0,0){0}}
\put(220.0,857.0){\rule[-0.200pt]{0.400pt}{4.818pt}}
\put(394.0,113.0){\rule[-0.200pt]{0.400pt}{4.818pt}}
\put(394,68){\makebox(0,0){5}}
\put(394.0,857.0){\rule[-0.200pt]{0.400pt}{4.818pt}}
\put(567.0,113.0){\rule[-0.200pt]{0.400pt}{4.818pt}}
\put(567,68){\makebox(0,0){10}}
\put(567.0,857.0){\rule[-0.200pt]{0.400pt}{4.818pt}}
\put(741.0,113.0){\rule[-0.200pt]{0.400pt}{4.818pt}}
\put(741,68){\makebox(0,0){15}}
\put(741.0,857.0){\rule[-0.200pt]{0.400pt}{4.818pt}}
\put(915.0,113.0){\rule[-0.200pt]{0.400pt}{4.818pt}}
\put(915,68){\makebox(0,0){20}}
\put(915.0,857.0){\rule[-0.200pt]{0.400pt}{4.818pt}}
\put(1089.0,113.0){\rule[-0.200pt]{0.400pt}{4.818pt}}
\put(1089,68){\makebox(0,0){25}}
\put(1089.0,857.0){\rule[-0.200pt]{0.400pt}{4.818pt}}
\put(1262.0,113.0){\rule[-0.200pt]{0.400pt}{4.818pt}}
\put(1262,68){\makebox(0,0){30}}
\put(1262.0,857.0){\rule[-0.200pt]{0.400pt}{4.818pt}}
\put(1436.0,113.0){\rule[-0.200pt]{0.400pt}{4.818pt}}
\put(1436,68){\makebox(0,0){35}}
\put(1436.0,857.0){\rule[-0.200pt]{0.400pt}{4.818pt}}
\put(220.0,113.0){\rule[-0.200pt]{292.934pt}{0.400pt}}
\put(1436.0,113.0){\rule[-0.200pt]{0.400pt}{184.048pt}}
\put(220.0,877.0){\rule[-0.200pt]{292.934pt}{0.400pt}}
\put(45,495){\makebox(0,0){Error}}
\put(828,23){\makebox(0,0){$r$}}
\put(220.0,113.0){\rule[-0.200pt]{0.400pt}{184.048pt}}
\put(1306,812){\makebox(0,0)[r]{'second'}}
\put(1328.0,812.0){\rule[-0.200pt]{15.899pt}{0.400pt}}
\put(289,805){\usebox{\plotpoint}}
\multiput(289.58,796.02)(0.499,-2.588){137}{\rule{0.120pt}{2.163pt}}
\multiput(288.17,800.51)(70.000,-356.511){2}{\rule{0.400pt}{1.081pt}}
\multiput(359.58,438.24)(0.499,-1.614){135}{\rule{0.120pt}{1.387pt}}
\multiput(358.17,441.12)(69.000,-219.121){2}{\rule{0.400pt}{0.693pt}}
\multiput(428.00,220.92)(0.522,-0.499){131}{\rule{0.518pt}{0.120pt}}
\multiput(428.00,221.17)(68.925,-67.000){2}{\rule{0.259pt}{0.400pt}}
\multiput(498.00,153.92)(1.584,-0.496){41}{\rule{1.355pt}{0.120pt}}
\multiput(498.00,154.17)(66.189,-22.000){2}{\rule{0.677pt}{0.400pt}}
\multiput(567.00,131.93)(4.048,-0.489){15}{\rule{3.211pt}{0.118pt}}
\multiput(567.00,132.17)(63.335,-9.000){2}{\rule{1.606pt}{0.400pt}}
\multiput(637.00,122.94)(9.986,-0.468){5}{\rule{7.000pt}{0.113pt}}
\multiput(637.00,123.17)(54.471,-4.000){2}{\rule{3.500pt}{0.400pt}}
\multiput(706.00,118.95)(15.421,-0.447){3}{\rule{9.433pt}{0.108pt}}
\multiput(706.00,119.17)(50.421,-3.000){2}{\rule{4.717pt}{0.400pt}}
\put(776,115.67){\rule{16.622pt}{0.400pt}}
\multiput(776.00,116.17)(34.500,-1.000){2}{\rule{8.311pt}{0.400pt}}
\put(845,114.67){\rule{16.863pt}{0.400pt}}
\multiput(845.00,115.17)(35.000,-1.000){2}{\rule{8.431pt}{0.400pt}}
\put(915,113.67){\rule{16.622pt}{0.400pt}}
\multiput(915.00,114.17)(34.500,-1.000){2}{\rule{8.311pt}{0.400pt}}
\put(1123,112.67){\rule{16.863pt}{0.400pt}}
\multiput(1123.00,113.17)(35.000,-1.000){2}{\rule{8.431pt}{0.400pt}}
\put(984.0,114.0){\rule[-0.200pt]{33.485pt}{0.400pt}}
\put(1350,812){\raisebox{-.8pt}{\makebox(0,0){$\Diamond$}}}
\put(289,805){\raisebox{-.8pt}{\makebox(0,0){$\Diamond$}}}
\put(359,444){\raisebox{-.8pt}{\makebox(0,0){$\Diamond$}}}
\put(428,222){\raisebox{-.8pt}{\makebox(0,0){$\Diamond$}}}
\put(498,155){\raisebox{-.8pt}{\makebox(0,0){$\Diamond$}}}
\put(567,133){\raisebox{-.8pt}{\makebox(0,0){$\Diamond$}}}
\put(637,124){\raisebox{-.8pt}{\makebox(0,0){$\Diamond$}}}
\put(706,120){\raisebox{-.8pt}{\makebox(0,0){$\Diamond$}}}
\put(776,117){\raisebox{-.8pt}{\makebox(0,0){$\Diamond$}}}
\put(845,116){\raisebox{-.8pt}{\makebox(0,0){$\Diamond$}}}
\put(915,115){\raisebox{-.8pt}{\makebox(0,0){$\Diamond$}}}
\put(984,114){\raisebox{-.8pt}{\makebox(0,0){$\Diamond$}}}
\put(1054,114){\raisebox{-.8pt}{\makebox(0,0){$\Diamond$}}}
\put(1123,114){\raisebox{-.8pt}{\makebox(0,0){$\Diamond$}}}
\put(1193,113){\raisebox{-.8pt}{\makebox(0,0){$\Diamond$}}}
\put(1262,113){\raisebox{-.8pt}{\makebox(0,0){$\Diamond$}}}
\put(1332,113){\raisebox{-.8pt}{\makebox(0,0){$\Diamond$}}}
\put(1193.0,113.0){\rule[-0.200pt]{33.485pt}{0.400pt}}
\put(1306,767){\makebox(0,0)[r]{'fourth'}}
\multiput(1328,767)(20.756,0.000){4}{\usebox{\plotpoint}}
\put(1394,767){\usebox{\plotpoint}}
\put(289,468){\usebox{\plotpoint}}
\multiput(289,468)(6.191,-19.811){12}{\usebox{\plotpoint}}
\multiput(359,244)(10.958,-17.627){6}{\usebox{\plotpoint}}
\multiput(428,133)(20.234,-4.625){4}{\usebox{\plotpoint}}
\multiput(498,117)(20.736,-0.902){3}{\usebox{\plotpoint}}
\multiput(567,114)(20.756,0.000){3}{\usebox{\plotpoint}}
\multiput(637,114)(20.753,-0.301){4}{\usebox{\plotpoint}}
\multiput(706,113)(20.756,0.000){3}{\usebox{\plotpoint}}
\multiput(776,113)(20.756,0.000){3}{\usebox{\plotpoint}}
\multiput(845,113)(20.756,0.000){4}{\usebox{\plotpoint}}
\multiput(915,113)(20.756,0.000){3}{\usebox{\plotpoint}}
\multiput(984,113)(20.756,0.000){3}{\usebox{\plotpoint}}
\multiput(1054,113)(20.756,0.000){4}{\usebox{\plotpoint}}
\multiput(1123,113)(20.756,0.000){3}{\usebox{\plotpoint}}
\multiput(1193,113)(20.756,0.000){3}{\usebox{\plotpoint}}
\multiput(1262,113)(20.756,0.000){4}{\usebox{\plotpoint}}
\multiput(1332,113)(20.756,0.000){3}{\usebox{\plotpoint}}
\put(1401,113){\usebox{\plotpoint}}
\put(1350,767){\makebox(0,0){$+$}}
\put(289,468){\makebox(0,0){$+$}}
\put(359,244){\makebox(0,0){$+$}}
\put(428,133){\makebox(0,0){$+$}}
\put(498,117){\makebox(0,0){$+$}}
\put(567,114){\makebox(0,0){$+$}}
\put(637,114){\makebox(0,0){$+$}}
\put(706,113){\makebox(0,0){$+$}}
\put(776,113){\makebox(0,0){$+$}}
\put(845,113){\makebox(0,0){$+$}}
\put(915,113){\makebox(0,0){$+$}}
\put(984,113){\makebox(0,0){$+$}}
\put(1054,113){\makebox(0,0){$+$}}
\put(1123,113){\makebox(0,0){$+$}}
\put(1193,113){\makebox(0,0){$+$}}
\put(1262,113){\makebox(0,0){$+$}}
\put(1332,113){\makebox(0,0){$+$}}
\put(1401,113){\makebox(0,0){$+$}}
\end{picture}
\end{center}
\caption{ Uniform domain calculations illustrating 
the improved accuracy of the fourth order
approximation over the second order case.}
\end{figure}

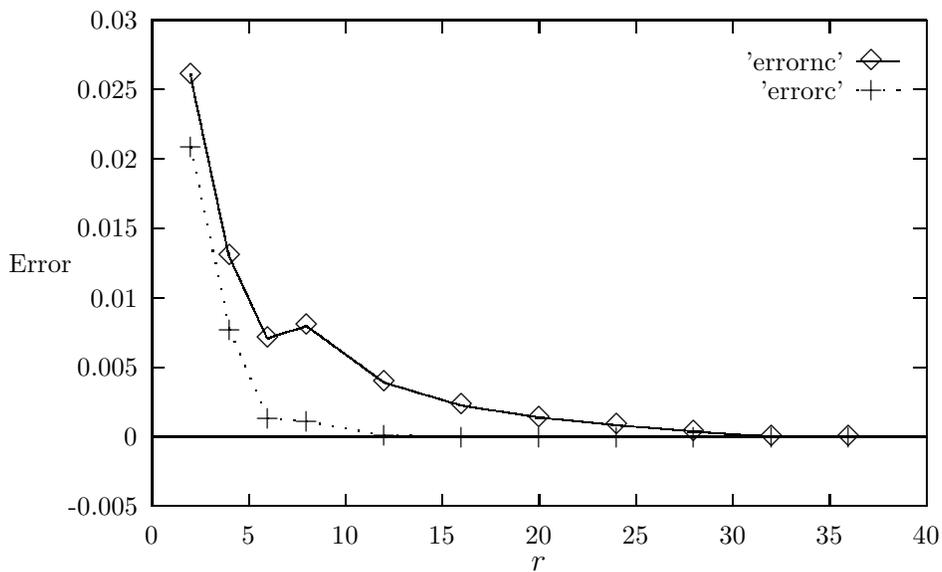
\begin{figure}
\begin{center}
\setlength{\unitlength}{0.240900pt}
\ifx\plotpoint\undefined\newsavebox{\plotpoint}\fi
\sbox{\plotpoint}{\rule[-0.200pt]{0.400pt}{0.400pt}}%
\begin{picture}(1500,900)(0,0)
\font\gnuplot=cmr10 at 10pt
\gnuplot
\sbox{\plotpoint}{\rule[-0.200pt]{0.400pt}{0.400pt}}%
\put(220.0,222.0){\rule[-0.200pt]{292.934pt}{0.400pt}}
\put(220.0,113.0){\rule[-0.200pt]{0.400pt}{184.048pt}}
\put(220.0,113.0){\rule[-0.200pt]{4.818pt}{0.400pt}}
\put(198,113){\makebox(0,0)[r]{-0.005}}
\put(1416.0,113.0){\rule[-0.200pt]{4.818pt}{0.400pt}}
\put(220.0,222.0){\rule[-0.200pt]{4.818pt}{0.400pt}}
\put(198,222){\makebox(0,0)[r]{0}}
\put(1416.0,222.0){\rule[-0.200pt]{4.818pt}{0.400pt}}
\put(220.0,331.0){\rule[-0.200pt]{4.818pt}{0.400pt}}
\put(198,331){\makebox(0,0)[r]{0.005}}
\put(1416.0,331.0){\rule[-0.200pt]{4.818pt}{0.400pt}}
\put(220.0,440.0){\rule[-0.200pt]{4.818pt}{0.400pt}}
\put(198,440){\makebox(0,0)[r]{0.01}}
\put(1416.0,440.0){\rule[-0.200pt]{4.818pt}{0.400pt}}
\put(220.0,550.0){\rule[-0.200pt]{4.818pt}{0.400pt}}
\put(198,550){\makebox(0,0)[r]{0.015}}
\put(1416.0,550.0){\rule[-0.200pt]{4.818pt}{0.400pt}}
\put(220.0,659.0){\rule[-0.200pt]{4.818pt}{0.400pt}}
\put(198,659){\makebox(0,0)[r]{0.02}}
\put(1416.0,659.0){\rule[-0.200pt]{4.818pt}{0.400pt}}
\put(220.0,768.0){\rule[-0.200pt]{4.818pt}{0.400pt}}
\put(198,768){\makebox(0,0)[r]{0.025}}
\put(1416.0,768.0){\rule[-0.200pt]{4.818pt}{0.400pt}}
\put(220.0,877.0){\rule[-0.200pt]{4.818pt}{0.400pt}}
\put(198,877){\makebox(0,0)[r]{0.03}}
\put(1416.0,877.0){\rule[-0.200pt]{4.818pt}{0.400pt}}
\put(220.0,113.0){\rule[-0.200pt]{0.400pt}{4.818pt}}
\put(220,68){\makebox(0,0){0}}
\put(220.0,857.0){\rule[-0.200pt]{0.400pt}{4.818pt}}
\put(372.0,113.0){\rule[-0.200pt]{0.400pt}{4.818pt}}
\put(372,68){\makebox(0,0){5}}
\put(372.0,857.0){\rule[-0.200pt]{0.400pt}{4.818pt}}
\put(524.0,113.0){\rule[-0.200pt]{0.400pt}{4.818pt}}
\put(524,68){\makebox(0,0){10}}
\put(524.0,857.0){\rule[-0.200pt]{0.400pt}{4.818pt}}
\put(676.0,113.0){\rule[-0.200pt]{0.400pt}{4.818pt}}
\put(676,68){\makebox(0,0){15}}
\put(676.0,857.0){\rule[-0.200pt]{0.400pt}{4.818pt}}
\put(828.0,113.0){\rule[-0.200pt]{0.400pt}{4.818pt}}
\put(828,68){\makebox(0,0){20}}
\put(828.0,857.0){\rule[-0.200pt]{0.400pt}{4.818pt}}
\put(980.0,113.0){\rule[-0.200pt]{0.400pt}{4.818pt}}
\put(980,68){\makebox(0,0){25}}
\put(980.0,857.0){\rule[-0.200pt]{0.400pt}{4.818pt}}
\put(1132.0,113.0){\rule[-0.200pt]{0.400pt}{4.818pt}}
\put(1132,68){\makebox(0,0){30}}
\put(1132.0,857.0){\rule[-0.200pt]{0.400pt}{4.818pt}}
\put(1284.0,113.0){\rule[-0.200pt]{0.400pt}{4.818pt}}
\put(1284,68){\makebox(0,0){35}}
\put(1284.0,857.0){\rule[-0.200pt]{0.400pt}{4.818pt}}
\put(1436.0,113.0){\rule[-0.200pt]{0.400pt}{4.818pt}}
\put(1436,68){\makebox(0,0){40}}
\put(1436.0,857.0){\rule[-0.200pt]{0.400pt}{4.818pt}}
\put(220.0,113.0){\rule[-0.200pt]{292.934pt}{0.400pt}}
\put(1436.0,113.0){\rule[-0.200pt]{0.400pt}{184.048pt}}
\put(220.0,877.0){\rule[-0.200pt]{292.934pt}{0.400pt}}
\put(45,495){\makebox(0,0){Error}}
\put(828,23){\makebox(0,0){$r$}}
\put(220.0,113.0){\rule[-0.200pt]{0.400pt}{184.048pt}}
\put(1306,812){\makebox(0,0)[r]{'errornc'}}
\put(1328.0,812.0){\rule[-0.200pt]{15.899pt}{0.400pt}}
\put(281,791){\usebox{\plotpoint}}
\multiput(281.58,782.83)(0.499,-2.346){119}{\rule{0.120pt}{1.969pt}}
\multiput(280.17,786.91)(61.000,-280.914){2}{\rule{0.400pt}{0.984pt}}
\multiput(342.58,502.01)(0.499,-1.078){117}{\rule{0.120pt}{0.960pt}}
\multiput(341.17,504.01)(60.000,-127.007){2}{\rule{0.400pt}{0.480pt}}
\multiput(402.00,377.58)(1.625,0.495){35}{\rule{1.384pt}{0.119pt}}
\multiput(402.00,376.17)(58.127,19.000){2}{\rule{0.692pt}{0.400pt}}
\multiput(463.00,394.92)(0.686,-0.499){175}{\rule{0.648pt}{0.120pt}}
\multiput(463.00,395.17)(120.654,-89.000){2}{\rule{0.324pt}{0.400pt}}
\multiput(585.00,305.92)(1.691,-0.498){69}{\rule{1.444pt}{0.120pt}}
\multiput(585.00,306.17)(118.002,-36.000){2}{\rule{0.722pt}{0.400pt}}
\multiput(706.00,269.92)(3.262,-0.495){35}{\rule{2.668pt}{0.119pt}}
\multiput(706.00,270.17)(116.462,-19.000){2}{\rule{1.334pt}{0.400pt}}
\multiput(828.00,250.92)(5.235,-0.492){21}{\rule{4.167pt}{0.119pt}}
\multiput(828.00,251.17)(113.352,-12.000){2}{\rule{2.083pt}{0.400pt}}
\multiput(950.00,238.92)(6.281,-0.491){17}{\rule{4.940pt}{0.118pt}}
\multiput(950.00,239.17)(110.747,-10.000){2}{\rule{2.470pt}{0.400pt}}
\multiput(1071.00,228.93)(8.022,-0.488){13}{\rule{6.200pt}{0.117pt}}
\multiput(1071.00,229.17)(109.132,-8.000){2}{\rule{3.100pt}{0.400pt}}
\put(1350,812){\raisebox{-.8pt}{\makebox(0,0){$\Diamond$}}}
\put(281,791){\raisebox{-.8pt}{\makebox(0,0){$\Diamond$}}}
\put(342,506){\raisebox{-.8pt}{\makebox(0,0){$\Diamond$}}}
\put(402,377){\raisebox{-.8pt}{\makebox(0,0){$\Diamond$}}}
\put(463,396){\raisebox{-.8pt}{\makebox(0,0){$\Diamond$}}}
\put(585,307){\raisebox{-.8pt}{\makebox(0,0){$\Diamond$}}}
\put(706,271){\raisebox{-.8pt}{\makebox(0,0){$\Diamond$}}}
\put(828,252){\raisebox{-.8pt}{\makebox(0,0){$\Diamond$}}}
\put(950,240){\raisebox{-.8pt}{\makebox(0,0){$\Diamond$}}}
\put(1071,230){\raisebox{-.8pt}{\makebox(0,0){$\Diamond$}}}
\put(1193,222){\raisebox{-.8pt}{\makebox(0,0){$\Diamond$}}}
\put(1314,222){\raisebox{-.8pt}{\makebox(0,0){$\Diamond$}}}
\put(1193.0,222.0){\rule[-0.200pt]{29.149pt}{0.400pt}}
\put(1306,767){\makebox(0,0)[r]{'errorc'}}
\multiput(1328,767)(20.756,0.000){4}{\usebox{\plotpoint}}
\put(1394,767){\usebox{\plotpoint}}
\put(281,677){\usebox{\plotpoint}}
\multiput(281,677)(4.315,-20.302){15}{\usebox{\plotpoint}}
\multiput(342,390)(8.226,-19.056){7}{\usebox{\plotpoint}}
\multiput(402,251)(20.686,-1.696){3}{\usebox{\plotpoint}}
\multiput(463,246)(20.426,-3.683){6}{\usebox{\plotpoint}}
\multiput(585,224)(20.753,-0.343){6}{\usebox{\plotpoint}}
\multiput(706,222)(20.756,0.000){6}{\usebox{\plotpoint}}
\multiput(828,222)(20.756,0.000){5}{\usebox{\plotpoint}}
\multiput(950,222)(20.756,0.000){6}{\usebox{\plotpoint}}
\multiput(1071,222)(20.756,0.000){6}{\usebox{\plotpoint}}
\multiput(1193,222)(20.756,0.000){6}{\usebox{\plotpoint}}
\put(1314,222){\usebox{\plotpoint}}
\put(1350,767){\makebox(0,0){$+$}}
\put(281,677){\makebox(0,0){$+$}}
\put(342,390){\makebox(0,0){$+$}}
\put(402,251){\makebox(0,0){$+$}}
\put(463,246){\makebox(0,0){$+$}}
\put(585,224){\makebox(0,0){$+$}}
\put(706,222){\makebox(0,0){$+$}}
\put(828,222){\makebox(0,0){$+$}}
\put(950,222){\makebox(0,0){$+$}}
\put(1071,222){\makebox(0,0){$+$}}
\put(1193,222){\makebox(0,0){$+$}}
\put(1314,222){\makebox(0,0){$+$}}
\end{picture}
\end{center}
\caption{Impact of nonconservation on the accuracy. The refinement
patch edge is at $r=8$. The diamonds are for the nonconservative
calculation and the crosses for the conservative case.}
\end{figure}

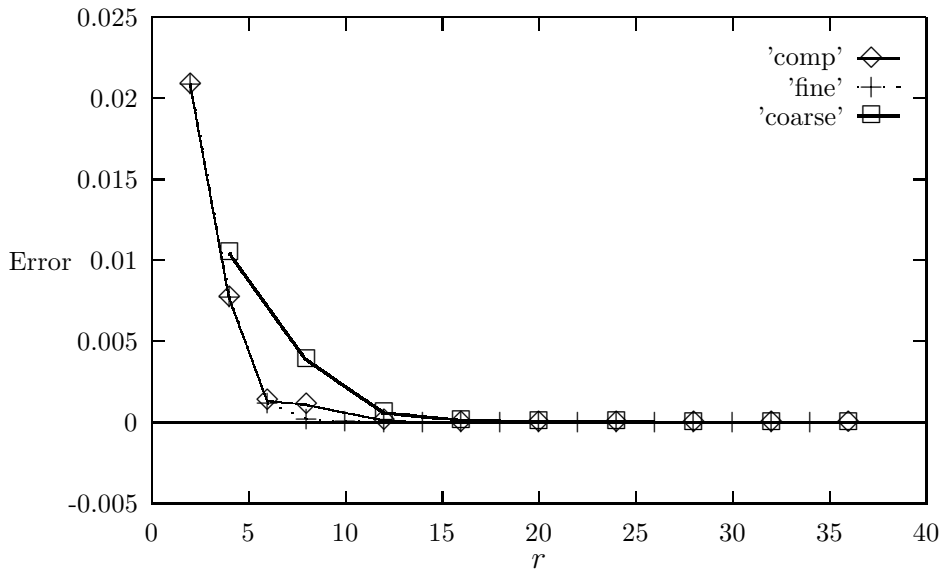
\begin{figure}
\begin{center}
\setlength{\unitlength}{0.240900pt}
\ifx\plotpoint\undefined\newsavebox{\plotpoint}\fi
\sbox{\plotpoint}{\rule[-0.200pt]{0.400pt}{0.400pt}}%
\begin{picture}(1500,900)(0,0)
\font\gnuplot=cmr10 at 10pt
\gnuplot
\sbox{\plotpoint}{\rule[-0.200pt]{0.400pt}{0.400pt}}%
\put(220.0,240.0){\rule[-0.200pt]{292.934pt}{0.400pt}}
\put(220.0,113.0){\rule[-0.200pt]{0.400pt}{184.048pt}}
\put(220.0,113.0){\rule[-0.200pt]{4.818pt}{0.400pt}}
\put(198,113){\makebox(0,0)[r]{-0.005}}
\put(1416.0,113.0){\rule[-0.200pt]{4.818pt}{0.400pt}}
\put(220.0,240.0){\rule[-0.200pt]{4.818pt}{0.400pt}}
\put(198,240){\makebox(0,0)[r]{0}}
\put(1416.0,240.0){\rule[-0.200pt]{4.818pt}{0.400pt}}
\put(220.0,368.0){\rule[-0.200pt]{4.818pt}{0.400pt}}
\put(198,368){\makebox(0,0)[r]{0.005}}
\put(1416.0,368.0){\rule[-0.200pt]{4.818pt}{0.400pt}}
\put(220.0,495.0){\rule[-0.200pt]{4.818pt}{0.400pt}}
\put(198,495){\makebox(0,0)[r]{0.01}}
\put(1416.0,495.0){\rule[-0.200pt]{4.818pt}{0.400pt}}
\put(220.0,622.0){\rule[-0.200pt]{4.818pt}{0.400pt}}
\put(198,622){\makebox(0,0)[r]{0.015}}
\put(1416.0,622.0){\rule[-0.200pt]{4.818pt}{0.400pt}}
\put(220.0,750.0){\rule[-0.200pt]{4.818pt}{0.400pt}}
\put(198,750){\makebox(0,0)[r]{0.02}}
\put(1416.0,750.0){\rule[-0.200pt]{4.818pt}{0.400pt}}
\put(220.0,877.0){\rule[-0.200pt]{4.818pt}{0.400pt}}
\put(198,877){\makebox(0,0)[r]{0.025}}
\put(1416.0,877.0){\rule[-0.200pt]{4.818pt}{0.400pt}}
\put(220.0,113.0){\rule[-0.200pt]{0.400pt}{4.818pt}}
\put(220,68){\makebox(0,0){0}}
\put(220.0,857.0){\rule[-0.200pt]{0.400pt}{4.818pt}}
\put(372.0,113.0){\rule[-0.200pt]{0.400pt}{4.818pt}}
\put(372,68){\makebox(0,0){5}}
\put(372.0,857.0){\rule[-0.200pt]{0.400pt}{4.818pt}}
\put(524.0,113.0){\rule[-0.200pt]{0.400pt}{4.818pt}}
\put(524,68){\makebox(0,0){10}}
\put(524.0,857.0){\rule[-0.200pt]{0.400pt}{4.818pt}}
\put(676.0,113.0){\rule[-0.200pt]{0.400pt}{4.818pt}}
\put(676,68){\makebox(0,0){15}}
\put(676.0,857.0){\rule[-0.200pt]{0.400pt}{4.818pt}}
\put(828.0,113.0){\rule[-0.200pt]{0.400pt}{4.818pt}}
\put(828,68){\makebox(0,0){20}}
\put(828.0,857.0){\rule[-0.200pt]{0.400pt}{4.818pt}}
\put(980.0,113.0){\rule[-0.200pt]{0.400pt}{4.818pt}}
\put(980,68){\makebox(0,0){25}}
\put(980.0,857.0){\rule[-0.200pt]{0.400pt}{4.818pt}}
\put(1132.0,113.0){\rule[-0.200pt]{0.400pt}{4.818pt}}
\put(1132,68){\makebox(0,0){30}}
\put(1132.0,857.0){\rule[-0.200pt]{0.400pt}{4.818pt}}
\put(1284.0,113.0){\rule[-0.200pt]{0.400pt}{4.818pt}}
\put(1284,68){\makebox(0,0){35}}
\put(1284.0,857.0){\rule[-0.200pt]{0.400pt}{4.818pt}}
\put(1436.0,113.0){\rule[-0.200pt]{0.400pt}{4.818pt}}
\put(1436,68){\makebox(0,0){40}}
\put(1436.0,857.0){\rule[-0.200pt]{0.400pt}{4.818pt}}
\put(220.0,113.0){\rule[-0.200pt]{292.934pt}{0.400pt}}
\put(1436.0,113.0){\rule[-0.200pt]{0.400pt}{184.048pt}}
\put(220.0,877.0){\rule[-0.200pt]{292.934pt}{0.400pt}}
\put(45,495){\makebox(0,0){Error}}
\put(828,23){\makebox(0,0){$r$}}
\put(220.0,113.0){\rule[-0.200pt]{0.400pt}{184.048pt}}
\put(1306,812){\makebox(0,0)[r]{'comp'}}
\put(1328.0,812.0){\rule[-0.200pt]{15.899pt}{0.400pt}}
\put(281,771){\usebox{\plotpoint}}
\multiput(281.58,761.47)(0.499,-2.758){119}{\rule{0.120pt}{2.297pt}}
\multiput(280.17,766.23)(61.000,-330.233){2}{\rule{0.400pt}{1.148pt}}
\multiput(342.58,431.10)(0.499,-1.354){117}{\rule{0.120pt}{1.180pt}}
\multiput(341.17,433.55)(60.000,-159.551){2}{\rule{0.400pt}{0.590pt}}
\multiput(402.00,272.93)(5.463,-0.482){9}{\rule{4.167pt}{0.116pt}}
\multiput(402.00,273.17)(52.352,-6.000){2}{\rule{2.083pt}{0.400pt}}
\multiput(463.00,266.92)(2.467,-0.497){47}{\rule{2.052pt}{0.120pt}}
\multiput(463.00,267.17)(117.741,-25.000){2}{\rule{1.026pt}{0.400pt}}
\multiput(585.00,241.95)(26.807,-0.447){3}{\rule{16.233pt}{0.108pt}}
\multiput(585.00,242.17)(87.307,-3.000){2}{\rule{8.117pt}{0.400pt}}
\put(1350,812){\raisebox{-.8pt}{\makebox(0,0){$\Diamond$}}}
\put(281,771){\raisebox{-.8pt}{\makebox(0,0){$\Diamond$}}}
\put(342,436){\raisebox{-.8pt}{\makebox(0,0){$\Diamond$}}}
\put(402,274){\raisebox{-.8pt}{\makebox(0,0){$\Diamond$}}}
\put(463,268){\raisebox{-.8pt}{\makebox(0,0){$\Diamond$}}}
\put(585,243){\raisebox{-.8pt}{\makebox(0,0){$\Diamond$}}}
\put(706,240){\raisebox{-.8pt}{\makebox(0,0){$\Diamond$}}}
\put(828,240){\raisebox{-.8pt}{\makebox(0,0){$\Diamond$}}}
\put(950,240){\raisebox{-.8pt}{\makebox(0,0){$\Diamond$}}}
\put(1071,240){\raisebox{-.8pt}{\makebox(0,0){$\Diamond$}}}
\put(1193,240){\raisebox{-.8pt}{\makebox(0,0){$\Diamond$}}}
\put(1314,240){\raisebox{-.8pt}{\makebox(0,0){$\Diamond$}}}
\put(706.0,240.0){\rule[-0.200pt]{146.467pt}{0.400pt}}
\put(1306,767){\makebox(0,0)[r]{'fine'}}
\multiput(1328,767)(20.756,0.000){4}{\usebox{\plotpoint}}
\put(1394,767){\usebox{\plotpoint}}
\put(281,772){\usebox{\plotpoint}}
\multiput(281,772)(3.718,-20.420){17}{\usebox{\plotpoint}}
\multiput(342,437)(7.018,-19.533){8}{\usebox{\plotpoint}}
\multiput(402,270)(19.314,-7.599){4}{\usebox{\plotpoint}}
\multiput(463,246)(20.711,-1.358){3}{\usebox{\plotpoint}}
\multiput(524,242)(20.753,-0.340){2}{\usebox{\plotpoint}}
\multiput(585,241)(20.756,0.000){3}{\usebox{\plotpoint}}
\multiput(646,241)(20.756,0.000){3}{\usebox{\plotpoint}}
\multiput(706,241)(20.753,-0.340){3}{\usebox{\plotpoint}}
\multiput(767,240)(20.756,0.000){3}{\usebox{\plotpoint}}
\multiput(828,240)(20.756,0.000){3}{\usebox{\plotpoint}}
\multiput(889,240)(20.756,0.000){3}{\usebox{\plotpoint}}
\multiput(950,240)(20.756,0.000){3}{\usebox{\plotpoint}}
\multiput(1010,240)(20.756,0.000){3}{\usebox{\plotpoint}}
\multiput(1071,240)(20.756,0.000){3}{\usebox{\plotpoint}}
\multiput(1132,240)(20.756,0.000){3}{\usebox{\plotpoint}}
\multiput(1193,240)(20.756,0.000){3}{\usebox{\plotpoint}}
\put(1254,240){\usebox{\plotpoint}}
\put(1350,767){\makebox(0,0){$+$}}
\put(281,772){\makebox(0,0){$+$}}
\put(342,437){\makebox(0,0){$+$}}
\put(402,270){\makebox(0,0){$+$}}
\put(463,246){\makebox(0,0){$+$}}
\put(524,242){\makebox(0,0){$+$}}
\put(585,241){\makebox(0,0){$+$}}
\put(646,241){\makebox(0,0){$+$}}
\put(706,241){\makebox(0,0){$+$}}
\put(767,240){\makebox(0,0){$+$}}
\put(828,240){\makebox(0,0){$+$}}
\put(889,240){\makebox(0,0){$+$}}
\put(950,240){\makebox(0,0){$+$}}
\put(1010,240){\makebox(0,0){$+$}}
\put(1071,240){\makebox(0,0){$+$}}
\put(1132,240){\makebox(0,0){$+$}}
\put(1193,240){\makebox(0,0){$+$}}
\put(1254,240){\makebox(0,0){$+$}}
\sbox{\plotpoint}{\rule[-0.400pt]{0.800pt}{0.800pt}}%
\put(1306,722){\makebox(0,0)[r]{'coarse'}}
\put(1328.0,722.0){\rule[-0.400pt]{15.899pt}{0.800pt}}
\put(342,506){\usebox{\plotpoint}}
\multiput(343.41,500.56)(0.501,-0.695){235}{\rule{0.121pt}{1.311pt}}
\multiput(340.34,503.28)(121.000,-165.279){2}{\rule{0.800pt}{0.655pt}}
\multiput(463.00,336.09)(0.736,-0.501){159}{\rule{1.376pt}{0.121pt}}
\multiput(463.00,336.34)(119.144,-83.000){2}{\rule{0.688pt}{0.800pt}}
\multiput(585.00,253.08)(5.418,-0.511){17}{\rule{8.267pt}{0.123pt}}
\multiput(585.00,253.34)(103.842,-12.000){2}{\rule{4.133pt}{0.800pt}}
\put(706,240.34){\rule{29.390pt}{0.800pt}}
\multiput(706.00,241.34)(61.000,-2.000){2}{\rule{14.695pt}{0.800pt}}
\put(950,238.84){\rule{29.149pt}{0.800pt}}
\multiput(950.00,239.34)(60.500,-1.000){2}{\rule{14.574pt}{0.800pt}}
\put(828.0,241.0){\rule[-0.400pt]{29.390pt}{0.800pt}}
\put(1350,722){\raisebox{-.8pt}{\makebox(0,0){$\Box$}}}
\put(342,506){\raisebox{-.8pt}{\makebox(0,0){$\Box$}}}
\put(463,338){\raisebox{-.8pt}{\makebox(0,0){$\Box$}}}
\put(585,255){\raisebox{-.8pt}{\makebox(0,0){$\Box$}}}
\put(706,243){\raisebox{-.8pt}{\makebox(0,0){$\Box$}}}
\put(828,241){\raisebox{-.8pt}{\makebox(0,0){$\Box$}}}
\put(950,241){\raisebox{-.8pt}{\makebox(0,0){$\Box$}}}
\put(1071,240){\raisebox{-.8pt}{\makebox(0,0){$\Box$}}}
\put(1193,240){\raisebox{-.8pt}{\makebox(0,0){$\Box$}}}
\put(1314,240){\raisebox{-.8pt}{\makebox(0,0){$\Box$}}}
\put(1071.0,240.0){\rule[-0.400pt]{58.539pt}{0.800pt}}
\end{picture}
\end{center}
\caption{Comparison of the accuracy for the composite grid
with that on the uniform fine and coarse scales. The refinement
patch edge is at $r=8$. The diamonds are for the composite
grid, the crosses for the uniform fine grid, and the squares
for the uniform coarse grid. Note the potential value 
at the $r=8$ point on the 
composite grid is set by the coarse grid, which leads to 
the larger error there. The interior patch points are at
$r=2,4,6$.}
\end{figure}

\begin{figure}
\begin{center}
\setlength{\unitlength}{0.240900pt}
\ifx\plotpoint\undefined\newsavebox{\plotpoint}\fi
\sbox{\plotpoint}{\rule[-0.200pt]{0.400pt}{0.400pt}}%
\begin{picture}(1500,900)(0,0)
\font\gnuplot=cmr10 at 10pt
\gnuplot
\sbox{\plotpoint}{\rule[-0.200pt]{0.400pt}{0.400pt}}%
\put(220.0,198.0){\rule[-0.200pt]{292.934pt}{0.400pt}}
\put(220.0,113.0){\rule[-0.200pt]{4.818pt}{0.400pt}}
\put(198,113){\makebox(0,0)[r]{-0.0005}}
\put(1416.0,113.0){\rule[-0.200pt]{4.818pt}{0.400pt}}
\put(220.0,198.0){\rule[-0.200pt]{4.818pt}{0.400pt}}
\put(198,198){\makebox(0,0)[r]{0}}
\put(1416.0,198.0){\rule[-0.200pt]{4.818pt}{0.400pt}}
\put(220.0,283.0){\rule[-0.200pt]{4.818pt}{0.400pt}}
\put(198,283){\makebox(0,0)[r]{0.0005}}
\put(1416.0,283.0){\rule[-0.200pt]{4.818pt}{0.400pt}}
\put(220.0,368.0){\rule[-0.200pt]{4.818pt}{0.400pt}}
\put(198,368){\makebox(0,0)[r]{0.001}}
\put(1416.0,368.0){\rule[-0.200pt]{4.818pt}{0.400pt}}
\put(220.0,453.0){\rule[-0.200pt]{4.818pt}{0.400pt}}
\put(198,453){\makebox(0,0)[r]{0.0015}}
\put(1416.0,453.0){\rule[-0.200pt]{4.818pt}{0.400pt}}
\put(220.0,537.0){\rule[-0.200pt]{4.818pt}{0.400pt}}
\put(198,537){\makebox(0,0)[r]{0.002}}
\put(1416.0,537.0){\rule[-0.200pt]{4.818pt}{0.400pt}}
\put(220.0,622.0){\rule[-0.200pt]{4.818pt}{0.400pt}}
\put(198,622){\makebox(0,0)[r]{0.0025}}
\put(1416.0,622.0){\rule[-0.200pt]{4.818pt}{0.400pt}}
\put(220.0,707.0){\rule[-0.200pt]{4.818pt}{0.400pt}}
\put(198,707){\makebox(0,0)[r]{0.003}}
\put(1416.0,707.0){\rule[-0.200pt]{4.818pt}{0.400pt}}
\put(220.0,792.0){\rule[-0.200pt]{4.818pt}{0.400pt}}
\put(198,792){\makebox(0,0)[r]{0.0035}}
\put(1416.0,792.0){\rule[-0.200pt]{4.818pt}{0.400pt}}
\put(220.0,877.0){\rule[-0.200pt]{4.818pt}{0.400pt}}
\put(198,877){\makebox(0,0)[r]{0.004}}
\put(1416.0,877.0){\rule[-0.200pt]{4.818pt}{0.400pt}}
\put(220.0,113.0){\rule[-0.200pt]{0.400pt}{4.818pt}}
\put(220,68){\makebox(0,0){5}}
\put(220.0,857.0){\rule[-0.200pt]{0.400pt}{4.818pt}}
\put(394.0,113.0){\rule[-0.200pt]{0.400pt}{4.818pt}}
\put(394,68){\makebox(0,0){10}}
\put(394.0,857.0){\rule[-0.200pt]{0.400pt}{4.818pt}}
\put(567.0,113.0){\rule[-0.200pt]{0.400pt}{4.818pt}}
\put(567,68){\makebox(0,0){15}}
\put(567.0,857.0){\rule[-0.200pt]{0.400pt}{4.818pt}}
\put(741.0,113.0){\rule[-0.200pt]{0.400pt}{4.818pt}}
\put(741,68){\makebox(0,0){20}}
\put(741.0,857.0){\rule[-0.200pt]{0.400pt}{4.818pt}}
\put(915.0,113.0){\rule[-0.200pt]{0.400pt}{4.818pt}}
\put(915,68){\makebox(0,0){25}}
\put(915.0,857.0){\rule[-0.200pt]{0.400pt}{4.818pt}}
\put(1089.0,113.0){\rule[-0.200pt]{0.400pt}{4.818pt}}
\put(1089,68){\makebox(0,0){30}}
\put(1089.0,857.0){\rule[-0.200pt]{0.400pt}{4.818pt}}
\put(1262.0,113.0){\rule[-0.200pt]{0.400pt}{4.818pt}}
\put(1262,68){\makebox(0,0){35}}
\put(1262.0,857.0){\rule[-0.200pt]{0.400pt}{4.818pt}}
\put(1436.0,113.0){\rule[-0.200pt]{0.400pt}{4.818pt}}
\put(1436,68){\makebox(0,0){40}}
\put(1436.0,857.0){\rule[-0.200pt]{0.400pt}{4.818pt}}
\put(220.0,113.0){\rule[-0.200pt]{292.934pt}{0.400pt}}
\put(1436.0,113.0){\rule[-0.200pt]{0.400pt}{184.048pt}}
\put(220.0,877.0){\rule[-0.200pt]{292.934pt}{0.400pt}}
\put(45,495){\makebox(0,0){Error}}
\put(828,23){\makebox(0,0){$r$}}
\put(220.0,113.0){\rule[-0.200pt]{0.400pt}{184.048pt}}
\put(1306,812){\makebox(0,0)[r]{'compout'}}
\put(1328.0,812.0){\rule[-0.200pt]{15.899pt}{0.400pt}}
\put(324,383){\usebox{\plotpoint}}
\multiput(324.58,380.55)(0.499,-0.612){275}{\rule{0.120pt}{0.589pt}}
\multiput(323.17,381.78)(139.000,-168.777){2}{\rule{0.400pt}{0.295pt}}
\multiput(463.00,211.92)(4.738,-0.494){27}{\rule{3.807pt}{0.119pt}}
\multiput(463.00,212.17)(131.099,-15.000){2}{\rule{1.903pt}{0.400pt}}
\put(1350,812){\raisebox{-.8pt}{\makebox(0,0){$\Diamond$}}}
\put(324,383){\raisebox{-.8pt}{\makebox(0,0){$\Diamond$}}}
\put(463,213){\raisebox{-.8pt}{\makebox(0,0){$\Diamond$}}}
\put(602,198){\raisebox{-.8pt}{\makebox(0,0){$\Diamond$}}}
\put(741,198){\raisebox{-.8pt}{\makebox(0,0){$\Diamond$}}}
\put(880,198){\raisebox{-.8pt}{\makebox(0,0){$\Diamond$}}}
\put(1019,198){\raisebox{-.8pt}{\makebox(0,0){$\Diamond$}}}
\put(1158,198){\raisebox{-.8pt}{\makebox(0,0){$\Diamond$}}}
\put(1297,198){\raisebox{-.8pt}{\makebox(0,0){$\Diamond$}}}
\put(602.0,198.0){\rule[-0.200pt]{167.425pt}{0.400pt}}
\put(1306,767){\makebox(0,0)[r]{'coarse'}}
\multiput(1328,767)(20.756,0.000){4}{\usebox{\plotpoint}}
\put(1394,767){\usebox{\plotpoint}}
\put(324,852){\usebox{\plotpoint}}
\multiput(324,852)(5.051,-20.132){28}{\usebox{\plotpoint}}
\multiput(463,298)(17.933,-10.450){8}{\usebox{\plotpoint}}
\multiput(602,217)(20.665,-1.933){6}{\usebox{\plotpoint}}
\multiput(741,204)(20.747,-0.597){7}{\usebox{\plotpoint}}
\multiput(880,200)(20.755,-0.149){7}{\usebox{\plotpoint}}
\multiput(1019,199)(20.755,-0.149){7}{\usebox{\plotpoint}}
\multiput(1158,198)(20.756,0.000){6}{\usebox{\plotpoint}}
\put(1297,198){\usebox{\plotpoint}}
\put(1350,767){\makebox(0,0){$+$}}
\put(324,852){\makebox(0,0){$+$}}
\put(463,298){\makebox(0,0){$+$}}
\put(602,217){\makebox(0,0){$+$}}
\put(741,204){\makebox(0,0){$+$}}
\put(880,200){\makebox(0,0){$+$}}
\put(1019,199){\makebox(0,0){$+$}}
\put(1158,198){\makebox(0,0){$+$}}
\put(1297,198){\makebox(0,0){$+$}}
\end{picture}
\end{center}
\caption{Comparison of accuracy outside the refinement patch
with the uniform coarse grid results. The crosses are the 
uniform coarse grid results and the diamonds are for the 
composite grid computation.}
\end{figure}
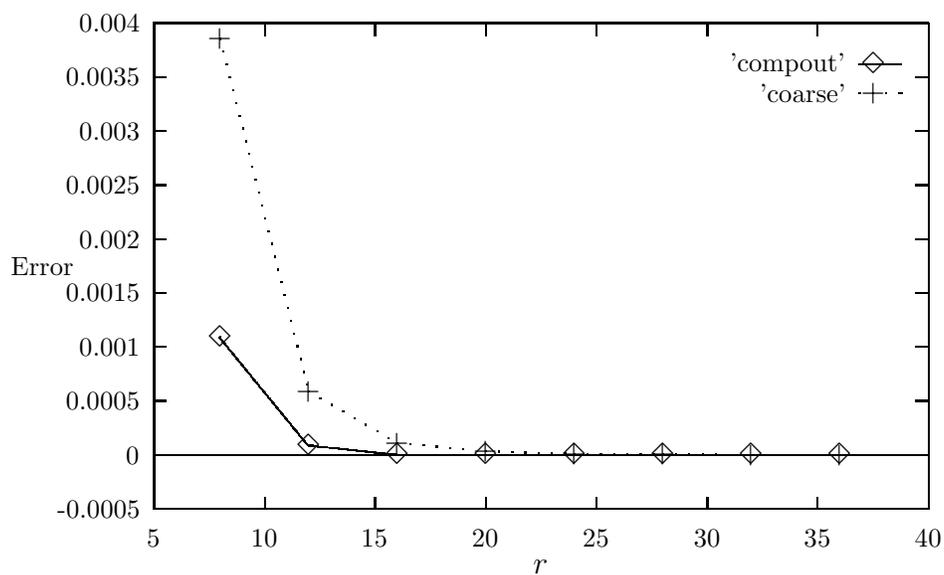

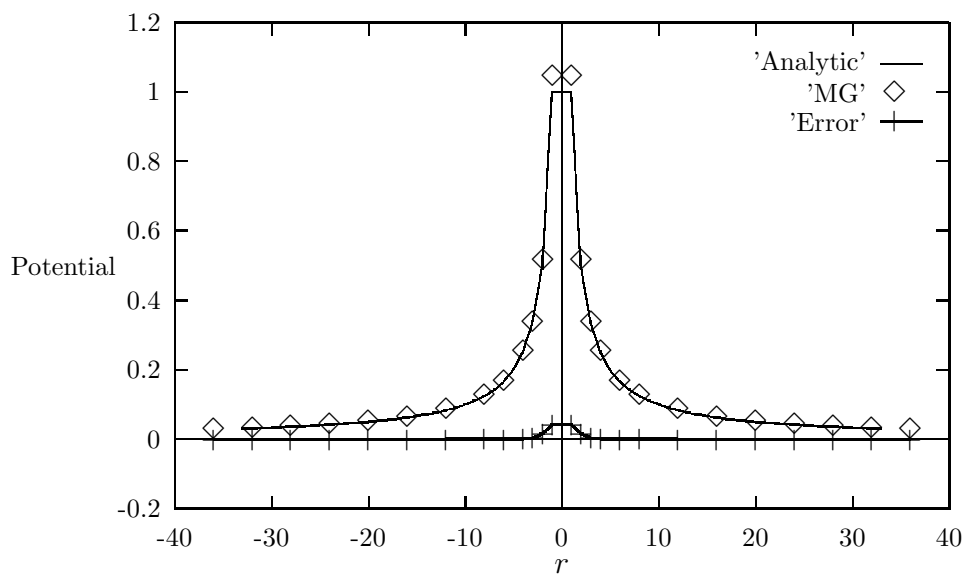
\begin{figure}
\begin{center}
\setlength{\unitlength}{0.240900pt}
\ifx\plotpoint\undefined\newsavebox{\plotpoint}\fi
\sbox{\plotpoint}{\rule[-0.200pt]{0.400pt}{0.400pt}}%
\begin{picture}(1500,900)(0,0)
\font\gnuplot=cmr10 at 10pt
\gnuplot
\sbox{\plotpoint}{\rule[-0.200pt]{0.400pt}{0.400pt}}%
\put(220.0,222.0){\rule[-0.200pt]{292.934pt}{0.400pt}}
\put(828.0,113.0){\rule[-0.200pt]{0.400pt}{184.048pt}}
\put(220.0,113.0){\rule[-0.200pt]{4.818pt}{0.400pt}}
\put(198,113){\makebox(0,0)[r]{-0.2}}
\put(1416.0,113.0){\rule[-0.200pt]{4.818pt}{0.400pt}}
\put(220.0,222.0){\rule[-0.200pt]{4.818pt}{0.400pt}}
\put(198,222){\makebox(0,0)[r]{0}}
\put(1416.0,222.0){\rule[-0.200pt]{4.818pt}{0.400pt}}
\put(220.0,331.0){\rule[-0.200pt]{4.818pt}{0.400pt}}
\put(198,331){\makebox(0,0)[r]{0.2}}
\put(1416.0,331.0){\rule[-0.200pt]{4.818pt}{0.400pt}}
\put(220.0,440.0){\rule[-0.200pt]{4.818pt}{0.400pt}}
\put(198,440){\makebox(0,0)[r]{0.4}}
\put(1416.0,440.0){\rule[-0.200pt]{4.818pt}{0.400pt}}
\put(220.0,550.0){\rule[-0.200pt]{4.818pt}{0.400pt}}
\put(198,550){\makebox(0,0)[r]{0.6}}
\put(1416.0,550.0){\rule[-0.200pt]{4.818pt}{0.400pt}}
\put(220.0,659.0){\rule[-0.200pt]{4.818pt}{0.400pt}}
\put(198,659){\makebox(0,0)[r]{0.8}}
\put(1416.0,659.0){\rule[-0.200pt]{4.818pt}{0.400pt}}
\put(220.0,768.0){\rule[-0.200pt]{4.818pt}{0.400pt}}
\put(198,768){\makebox(0,0)[r]{1}}
\put(1416.0,768.0){\rule[-0.200pt]{4.818pt}{0.400pt}}
\put(220.0,877.0){\rule[-0.200pt]{4.818pt}{0.400pt}}
\put(198,877){\makebox(0,0)[r]{1.2}}
\put(1416.0,877.0){\rule[-0.200pt]{4.818pt}{0.400pt}}
\put(220.0,113.0){\rule[-0.200pt]{0.400pt}{4.818pt}}
\put(220,68){\makebox(0,0){-40}}
\put(220.0,857.0){\rule[-0.200pt]{0.400pt}{4.818pt}}
\put(372.0,113.0){\rule[-0.200pt]{0.400pt}{4.818pt}}
\put(372,68){\makebox(0,0){-30}}
\put(372.0,857.0){\rule[-0.200pt]{0.400pt}{4.818pt}}
\put(524.0,113.0){\rule[-0.200pt]{0.400pt}{4.818pt}}
\put(524,68){\makebox(0,0){-20}}
\put(524.0,857.0){\rule[-0.200pt]{0.400pt}{4.818pt}}
\put(676.0,113.0){\rule[-0.200pt]{0.400pt}{4.818pt}}
\put(676,68){\makebox(0,0){-10}}
\put(676.0,857.0){\rule[-0.200pt]{0.400pt}{4.818pt}}
\put(828.0,113.0){\rule[-0.200pt]{0.400pt}{4.818pt}}
\put(828,68){\makebox(0,0){0}}
\put(828.0,857.0){\rule[-0.200pt]{0.400pt}{4.818pt}}
\put(980.0,113.0){\rule[-0.200pt]{0.400pt}{4.818pt}}
\put(980,68){\makebox(0,0){10}}
\put(980.0,857.0){\rule[-0.200pt]{0.400pt}{4.818pt}}
\put(1132.0,113.0){\rule[-0.200pt]{0.400pt}{4.818pt}}
\put(1132,68){\makebox(0,0){20}}
\put(1132.0,857.0){\rule[-0.200pt]{0.400pt}{4.818pt}}
\put(1284.0,113.0){\rule[-0.200pt]{0.400pt}{4.818pt}}
\put(1284,68){\makebox(0,0){30}}
\put(1284.0,857.0){\rule[-0.200pt]{0.400pt}{4.818pt}}
\put(1436.0,113.0){\rule[-0.200pt]{0.400pt}{4.818pt}}
\put(1436,68){\makebox(0,0){40}}
\put(1436.0,857.0){\rule[-0.200pt]{0.400pt}{4.818pt}}
\put(220.0,113.0){\rule[-0.200pt]{292.934pt}{0.400pt}}
\put(1436.0,113.0){\rule[-0.200pt]{0.400pt}{184.048pt}}
\put(220.0,877.0){\rule[-0.200pt]{292.934pt}{0.400pt}}
\put(45,495){\makebox(0,0){Potential}}
\put(828,23){\makebox(0,0){$r$}}
\put(220.0,113.0){\rule[-0.200pt]{0.400pt}{184.048pt}}
\put(1306,812){\makebox(0,0)[r]{'Analytic'}}
\put(1328.0,812.0){\rule[-0.200pt]{15.899pt}{0.400pt}}
\put(326,239){\usebox{\plotpoint}}
\put(342,238.67){\rule{3.614pt}{0.400pt}}
\multiput(342.00,238.17)(7.500,1.000){2}{\rule{1.807pt}{0.400pt}}
\put(326.0,239.0){\rule[-0.200pt]{3.854pt}{0.400pt}}
\put(372,239.67){\rule{3.614pt}{0.400pt}}
\multiput(372.00,239.17)(7.500,1.000){2}{\rule{1.807pt}{0.400pt}}
\put(387,240.67){\rule{3.614pt}{0.400pt}}
\multiput(387.00,240.17)(7.500,1.000){2}{\rule{1.807pt}{0.400pt}}
\put(357.0,240.0){\rule[-0.200pt]{3.613pt}{0.400pt}}
\put(418,241.67){\rule{3.614pt}{0.400pt}}
\multiput(418.00,241.17)(7.500,1.000){2}{\rule{1.807pt}{0.400pt}}
\put(433,242.67){\rule{3.614pt}{0.400pt}}
\multiput(433.00,242.17)(7.500,1.000){2}{\rule{1.807pt}{0.400pt}}
\put(448,243.67){\rule{3.614pt}{0.400pt}}
\multiput(448.00,243.17)(7.500,1.000){2}{\rule{1.807pt}{0.400pt}}
\put(463,244.67){\rule{3.614pt}{0.400pt}}
\multiput(463.00,244.17)(7.500,1.000){2}{\rule{1.807pt}{0.400pt}}
\put(478,245.67){\rule{3.854pt}{0.400pt}}
\multiput(478.00,245.17)(8.000,1.000){2}{\rule{1.927pt}{0.400pt}}
\put(494,246.67){\rule{3.614pt}{0.400pt}}
\multiput(494.00,246.17)(7.500,1.000){2}{\rule{1.807pt}{0.400pt}}
\put(509,247.67){\rule{3.614pt}{0.400pt}}
\multiput(509.00,247.17)(7.500,1.000){2}{\rule{1.807pt}{0.400pt}}
\put(524,249.17){\rule{3.100pt}{0.400pt}}
\multiput(524.00,248.17)(8.566,2.000){2}{\rule{1.550pt}{0.400pt}}
\put(539,250.67){\rule{3.614pt}{0.400pt}}
\multiput(539.00,250.17)(7.500,1.000){2}{\rule{1.807pt}{0.400pt}}
\put(554,252.17){\rule{3.300pt}{0.400pt}}
\multiput(554.00,251.17)(9.151,2.000){2}{\rule{1.650pt}{0.400pt}}
\put(570,254.17){\rule{3.100pt}{0.400pt}}
\multiput(570.00,253.17)(8.566,2.000){2}{\rule{1.550pt}{0.400pt}}
\multiput(585.00,256.61)(3.141,0.447){3}{\rule{2.100pt}{0.108pt}}
\multiput(585.00,255.17)(10.641,3.000){2}{\rule{1.050pt}{0.400pt}}
\put(600,259.17){\rule{3.100pt}{0.400pt}}
\multiput(600.00,258.17)(8.566,2.000){2}{\rule{1.550pt}{0.400pt}}
\multiput(615.00,261.61)(3.141,0.447){3}{\rule{2.100pt}{0.108pt}}
\multiput(615.00,260.17)(10.641,3.000){2}{\rule{1.050pt}{0.400pt}}
\multiput(630.00,264.60)(2.236,0.468){5}{\rule{1.700pt}{0.113pt}}
\multiput(630.00,263.17)(12.472,4.000){2}{\rule{0.850pt}{0.400pt}}
\multiput(646.00,268.60)(2.090,0.468){5}{\rule{1.600pt}{0.113pt}}
\multiput(646.00,267.17)(11.679,4.000){2}{\rule{0.800pt}{0.400pt}}
\multiput(661.00,272.59)(1.601,0.477){7}{\rule{1.300pt}{0.115pt}}
\multiput(661.00,271.17)(12.302,5.000){2}{\rule{0.650pt}{0.400pt}}
\multiput(676.00,277.59)(1.304,0.482){9}{\rule{1.100pt}{0.116pt}}
\multiput(676.00,276.17)(12.717,6.000){2}{\rule{0.550pt}{0.400pt}}
\multiput(691.00,283.59)(1.103,0.485){11}{\rule{0.957pt}{0.117pt}}
\multiput(691.00,282.17)(13.013,7.000){2}{\rule{0.479pt}{0.400pt}}
\multiput(706.00,290.58)(0.808,0.491){17}{\rule{0.740pt}{0.118pt}}
\multiput(706.00,289.17)(14.464,10.000){2}{\rule{0.370pt}{0.400pt}}
\multiput(722.00,300.58)(0.576,0.493){23}{\rule{0.562pt}{0.119pt}}
\multiput(722.00,299.17)(13.834,13.000){2}{\rule{0.281pt}{0.400pt}}
\multiput(737.58,313.00)(0.494,0.600){27}{\rule{0.119pt}{0.580pt}}
\multiput(736.17,313.00)(15.000,16.796){2}{\rule{0.400pt}{0.290pt}}
\multiput(752.58,331.00)(0.494,0.942){27}{\rule{0.119pt}{0.847pt}}
\multiput(751.17,331.00)(15.000,26.243){2}{\rule{0.400pt}{0.423pt}}
\multiput(767.58,359.00)(0.494,1.523){27}{\rule{0.119pt}{1.300pt}}
\multiput(766.17,359.00)(15.000,42.302){2}{\rule{0.400pt}{0.650pt}}
\multiput(782.58,404.00)(0.494,2.898){29}{\rule{0.119pt}{2.375pt}}
\multiput(781.17,404.00)(16.000,86.071){2}{\rule{0.400pt}{1.188pt}}
\multiput(798.58,495.00)(0.494,9.322){27}{\rule{0.119pt}{7.380pt}}
\multiput(797.17,495.00)(15.000,257.682){2}{\rule{0.400pt}{3.690pt}}
\put(402.0,242.0){\rule[-0.200pt]{3.854pt}{0.400pt}}
\multiput(843.58,737.36)(0.494,-9.322){27}{\rule{0.119pt}{7.380pt}}
\multiput(842.17,752.68)(15.000,-257.682){2}{\rule{0.400pt}{3.690pt}}
\multiput(858.58,485.14)(0.494,-2.898){29}{\rule{0.119pt}{2.375pt}}
\multiput(857.17,490.07)(16.000,-86.071){2}{\rule{0.400pt}{1.188pt}}
\multiput(874.58,398.60)(0.494,-1.523){27}{\rule{0.119pt}{1.300pt}}
\multiput(873.17,401.30)(15.000,-42.302){2}{\rule{0.400pt}{0.650pt}}
\multiput(889.58,355.49)(0.494,-0.942){27}{\rule{0.119pt}{0.847pt}}
\multiput(888.17,357.24)(15.000,-26.243){2}{\rule{0.400pt}{0.423pt}}
\multiput(904.58,328.59)(0.494,-0.600){27}{\rule{0.119pt}{0.580pt}}
\multiput(903.17,329.80)(15.000,-16.796){2}{\rule{0.400pt}{0.290pt}}
\multiput(919.00,311.92)(0.576,-0.493){23}{\rule{0.562pt}{0.119pt}}
\multiput(919.00,312.17)(13.834,-13.000){2}{\rule{0.281pt}{0.400pt}}
\multiput(934.00,298.92)(0.808,-0.491){17}{\rule{0.740pt}{0.118pt}}
\multiput(934.00,299.17)(14.464,-10.000){2}{\rule{0.370pt}{0.400pt}}
\multiput(950.00,288.93)(1.103,-0.485){11}{\rule{0.957pt}{0.117pt}}
\multiput(950.00,289.17)(13.013,-7.000){2}{\rule{0.479pt}{0.400pt}}
\multiput(965.00,281.93)(1.304,-0.482){9}{\rule{1.100pt}{0.116pt}}
\multiput(965.00,282.17)(12.717,-6.000){2}{\rule{0.550pt}{0.400pt}}
\multiput(980.00,275.93)(1.601,-0.477){7}{\rule{1.300pt}{0.115pt}}
\multiput(980.00,276.17)(12.302,-5.000){2}{\rule{0.650pt}{0.400pt}}
\multiput(995.00,270.94)(2.090,-0.468){5}{\rule{1.600pt}{0.113pt}}
\multiput(995.00,271.17)(11.679,-4.000){2}{\rule{0.800pt}{0.400pt}}
\multiput(1010.00,266.94)(2.236,-0.468){5}{\rule{1.700pt}{0.113pt}}
\multiput(1010.00,267.17)(12.472,-4.000){2}{\rule{0.850pt}{0.400pt}}
\multiput(1026.00,262.95)(3.141,-0.447){3}{\rule{2.100pt}{0.108pt}}
\multiput(1026.00,263.17)(10.641,-3.000){2}{\rule{1.050pt}{0.400pt}}
\put(1041,259.17){\rule{3.100pt}{0.400pt}}
\multiput(1041.00,260.17)(8.566,-2.000){2}{\rule{1.550pt}{0.400pt}}
\multiput(1056.00,257.95)(3.141,-0.447){3}{\rule{2.100pt}{0.108pt}}
\multiput(1056.00,258.17)(10.641,-3.000){2}{\rule{1.050pt}{0.400pt}}
\put(1071,254.17){\rule{3.100pt}{0.400pt}}
\multiput(1071.00,255.17)(8.566,-2.000){2}{\rule{1.550pt}{0.400pt}}
\put(1086,252.17){\rule{3.300pt}{0.400pt}}
\multiput(1086.00,253.17)(9.151,-2.000){2}{\rule{1.650pt}{0.400pt}}
\put(1102,250.67){\rule{3.614pt}{0.400pt}}
\multiput(1102.00,251.17)(7.500,-1.000){2}{\rule{1.807pt}{0.400pt}}
\put(1117,249.17){\rule{3.100pt}{0.400pt}}
\multiput(1117.00,250.17)(8.566,-2.000){2}{\rule{1.550pt}{0.400pt}}
\put(1132,247.67){\rule{3.614pt}{0.400pt}}
\multiput(1132.00,248.17)(7.500,-1.000){2}{\rule{1.807pt}{0.400pt}}
\put(1147,246.67){\rule{3.614pt}{0.400pt}}
\multiput(1147.00,247.17)(7.500,-1.000){2}{\rule{1.807pt}{0.400pt}}
\put(1162,245.67){\rule{3.854pt}{0.400pt}}
\multiput(1162.00,246.17)(8.000,-1.000){2}{\rule{1.927pt}{0.400pt}}
\put(1178,244.67){\rule{3.614pt}{0.400pt}}
\multiput(1178.00,245.17)(7.500,-1.000){2}{\rule{1.807pt}{0.400pt}}
\put(1193,243.67){\rule{3.614pt}{0.400pt}}
\multiput(1193.00,244.17)(7.500,-1.000){2}{\rule{1.807pt}{0.400pt}}
\put(1208,242.67){\rule{3.614pt}{0.400pt}}
\multiput(1208.00,243.17)(7.500,-1.000){2}{\rule{1.807pt}{0.400pt}}
\put(1223,241.67){\rule{3.614pt}{0.400pt}}
\multiput(1223.00,242.17)(7.500,-1.000){2}{\rule{1.807pt}{0.400pt}}
\put(813.0,768.0){\rule[-0.200pt]{7.227pt}{0.400pt}}
\put(1254,240.67){\rule{3.614pt}{0.400pt}}
\multiput(1254.00,241.17)(7.500,-1.000){2}{\rule{1.807pt}{0.400pt}}
\put(1269,239.67){\rule{3.614pt}{0.400pt}}
\multiput(1269.00,240.17)(7.500,-1.000){2}{\rule{1.807pt}{0.400pt}}
\put(1238.0,242.0){\rule[-0.200pt]{3.854pt}{0.400pt}}
\put(1299,238.67){\rule{3.614pt}{0.400pt}}
\multiput(1299.00,239.17)(7.500,-1.000){2}{\rule{1.807pt}{0.400pt}}
\put(1284.0,240.0){\rule[-0.200pt]{3.613pt}{0.400pt}}
\put(1314.0,239.0){\rule[-0.200pt]{3.854pt}{0.400pt}}
\put(1306,767){\makebox(0,0)[r]{'MG'}}
\put(1350,767){\raisebox{-.8pt}{\makebox(0,0){$\Diamond$}}}
\put(281,237){\raisebox{-.8pt}{\makebox(0,0){$\Diamond$}}}
\put(342,239){\raisebox{-.8pt}{\makebox(0,0){$\Diamond$}}}
\put(402,242){\raisebox{-.8pt}{\makebox(0,0){$\Diamond$}}}
\put(463,245){\raisebox{-.8pt}{\makebox(0,0){$\Diamond$}}}
\put(524,249){\raisebox{-.8pt}{\makebox(0,0){$\Diamond$}}}
\put(585,256){\raisebox{-.8pt}{\makebox(0,0){$\Diamond$}}}
\put(646,268){\raisebox{-.8pt}{\makebox(0,0){$\Diamond$}}}
\put(706,291){\raisebox{-.8pt}{\makebox(0,0){$\Diamond$}}}
\put(737,313){\raisebox{-.8pt}{\makebox(0,0){$\Diamond$}}}
\put(767,360){\raisebox{-.8pt}{\makebox(0,0){$\Diamond$}}}
\put(782,405){\raisebox{-.8pt}{\makebox(0,0){$\Diamond$}}}
\put(798,503){\raisebox{-.8pt}{\makebox(0,0){$\Diamond$}}}
\put(813,791){\raisebox{-.8pt}{\makebox(0,0){$\Diamond$}}}
\put(843,791){\raisebox{-.8pt}{\makebox(0,0){$\Diamond$}}}
\put(858,503){\raisebox{-.8pt}{\makebox(0,0){$\Diamond$}}}
\put(874,405){\raisebox{-.8pt}{\makebox(0,0){$\Diamond$}}}
\put(889,360){\raisebox{-.8pt}{\makebox(0,0){$\Diamond$}}}
\put(919,313){\raisebox{-.8pt}{\makebox(0,0){$\Diamond$}}}
\put(950,291){\raisebox{-.8pt}{\makebox(0,0){$\Diamond$}}}
\put(1010,268){\raisebox{-.8pt}{\makebox(0,0){$\Diamond$}}}
\put(1071,256){\raisebox{-.8pt}{\makebox(0,0){$\Diamond$}}}
\put(1132,249){\raisebox{-.8pt}{\makebox(0,0){$\Diamond$}}}
\put(1193,245){\raisebox{-.8pt}{\makebox(0,0){$\Diamond$}}}
\put(1254,242){\raisebox{-.8pt}{\makebox(0,0){$\Diamond$}}}
\put(1314,239){\raisebox{-.8pt}{\makebox(0,0){$\Diamond$}}}
\put(1375,237){\raisebox{-.8pt}{\makebox(0,0){$\Diamond$}}}
\sbox{\plotpoint}{\rule[-0.400pt]{0.800pt}{0.800pt}}%
\put(1306,722){\makebox(0,0)[r]{'Error'}}
\put(1328.0,722.0){\rule[-0.400pt]{15.899pt}{0.800pt}}
\put(281,222){\usebox{\plotpoint}}
\put(646,220.84){\rule{14.454pt}{0.800pt}}
\multiput(646.00,220.34)(30.000,1.000){2}{\rule{7.227pt}{0.800pt}}
\put(706,220.84){\rule{7.468pt}{0.800pt}}
\multiput(706.00,221.34)(15.500,-1.000){2}{\rule{3.734pt}{0.800pt}}
\put(737,220.84){\rule{7.227pt}{0.800pt}}
\multiput(737.00,220.34)(15.000,1.000){2}{\rule{3.613pt}{0.800pt}}
\put(767,221.84){\rule{3.614pt}{0.800pt}}
\multiput(767.00,221.34)(7.500,1.000){2}{\rule{1.807pt}{0.800pt}}
\multiput(782.00,225.40)(1.263,0.526){7}{\rule{2.029pt}{0.127pt}}
\multiput(782.00,222.34)(11.790,7.000){2}{\rule{1.014pt}{0.800pt}}
\multiput(798.00,232.41)(0.531,0.509){21}{\rule{1.057pt}{0.123pt}}
\multiput(798.00,229.34)(12.806,14.000){2}{\rule{0.529pt}{0.800pt}}
\put(281.0,222.0){\rule[-0.400pt]{87.928pt}{0.800pt}}
\multiput(843.00,243.09)(0.531,-0.509){21}{\rule{1.057pt}{0.123pt}}
\multiput(843.00,243.34)(12.806,-14.000){2}{\rule{0.529pt}{0.800pt}}
\multiput(858.00,229.08)(1.263,-0.526){7}{\rule{2.029pt}{0.127pt}}
\multiput(858.00,229.34)(11.790,-7.000){2}{\rule{1.014pt}{0.800pt}}
\put(874,221.84){\rule{3.614pt}{0.800pt}}
\multiput(874.00,222.34)(7.500,-1.000){2}{\rule{1.807pt}{0.800pt}}
\put(889,220.84){\rule{7.227pt}{0.800pt}}
\multiput(889.00,221.34)(15.000,-1.000){2}{\rule{3.613pt}{0.800pt}}
\put(919,220.84){\rule{7.468pt}{0.800pt}}
\multiput(919.00,220.34)(15.500,1.000){2}{\rule{3.734pt}{0.800pt}}
\put(950,220.84){\rule{14.454pt}{0.800pt}}
\multiput(950.00,221.34)(30.000,-1.000){2}{\rule{7.227pt}{0.800pt}}
\put(813.0,245.0){\rule[-0.400pt]{7.227pt}{0.800pt}}
\put(1350,722){\makebox(0,0){$+$}}
\put(281,222){\makebox(0,0){$+$}}
\put(342,222){\makebox(0,0){$+$}}
\put(402,222){\makebox(0,0){$+$}}
\put(463,222){\makebox(0,0){$+$}}
\put(524,222){\makebox(0,0){$+$}}
\put(585,222){\makebox(0,0){$+$}}
\put(646,222){\makebox(0,0){$+$}}
\put(706,223){\makebox(0,0){$+$}}
\put(737,222){\makebox(0,0){$+$}}
\put(767,223){\makebox(0,0){$+$}}
\put(782,224){\makebox(0,0){$+$}}
\put(798,231){\makebox(0,0){$+$}}
\put(813,245){\makebox(0,0){$+$}}
\put(843,245){\makebox(0,0){$+$}}
\put(858,231){\makebox(0,0){$+$}}
\put(874,224){\makebox(0,0){$+$}}
\put(889,223){\makebox(0,0){$+$}}
\put(919,222){\makebox(0,0){$+$}}
\put(950,223){\makebox(0,0){$+$}}
\put(1010,222){\makebox(0,0){$+$}}
\put(1071,222){\makebox(0,0){$+$}}
\put(1132,222){\makebox(0,0){$+$}}
\put(1193,222){\makebox(0,0){$+$}}
\put(1254,222){\makebox(0,0){$+$}}
\put(1314,222){\makebox(0,0){$+$}}
\put(1375,222){\makebox(0,0){$+$}}
\put(1010.0,222.0){\rule[-0.400pt]{87.928pt}{0.800pt}}
\end{picture}
\end{center}
\caption{Plotted are the analytical 1/r potential and 
the numerical results from the conservative mesh refinement
multigrid computation. The two fine patches span the
ranges -8. to 8. and -4. to 4. 
The lower curve gives the magnitude
of the difference between the exact and numerical results,
illustrating the larger errors near the source singularity.}
\end{figure}



\newpage

\begin{thebibliography}{99}


\bibitem{wolynes}Z. Luthey-Schulten, B. E. Ramirez, and 
P. G. Wolynes, J. Phys.\ Chem.\ {\bf 99}, 2177 (1995). 

\bibitem{dill}K. A. Dill, S. Bromberg, K. Yue, K. M. Fiebig,
D. P. Yee, P. D. Thomas, and H. S. Chan, Protein Sci.\ 
{\bf 4}, 561 (1995).

\bibitem{honig}N. Ben-Tal, B. Honig, C. Miller, and 
S. McLaughlin, Biophys.\ J. {\bf 73}, 1717 (1997). 

\bibitem{friesner}R. A. Friesner, Ann.\ Rev.\ Phys.\
Chem.\ {\bf 42}, 341 (1991). 

\bibitem{michelsen}J. A. Michelsen, in {\it Multigrid 
Methods III}, eds.\ W. Hackbusch and U. Trottenberg 
(Birkh\"{a}user, Berlin, 1991), p. 301. See other articles
in this collection as well.  

\bibitem{brandt}
A. Brandt, Math.\ Comp.\ {\bf 31}, 333 (1977). 

\bibitem{briggs}
W. L. Briggs, {\it A Multigrid Tutorial}, (SIAM, Philadelphia, 1987).

\bibitem{bai/brandt}
D. Bai and A. Brandt, SIAM J. Sci.\ Stat.\ 
Comput.\ {\bf 8}, 109 (1987). 

\bibitem{abeval}A. Brandt, S. McCormick, and J. Ruge,
SIAM J. Sci.\ Stat.\ Comput.\ {\bf 4}, 244 (1983). 

\bibitem{gridgen}{\it Computational Fluid Dynamics}, ed.\
H. Deconinck (Von Karman Institute for Fluid Dynamics,
Rhode-Saint-Genese, Belgium, 1995); {\it Mathematical Aspects
of Numerical Grid Generation}, ed.\ J. E. Castillo (SIAM, 
Philadelphia, 1991); {\it Modeling, Mesh Generation, and
Adaptive Numerical Methods for Partial Differential Equations},
eds.\ I. Babusa, {\it et al.} (Springer-Verlag, New York, 1995).

\bibitem{gygi}
F. Gygi and G. Galli, Phys.\ Rev.\ B
{\bf 52}, R2229 (1995). 

\bibitem{bernholc}
E. L. Briggs, D. J. Sullivan, and J. Bernholc, Phys.\ Rev.\ B
{\bf 52}, R5471 (1995); {\it ibid.} {\bf 54}, 14362 (1996);
J. Bernholc, E. L. Briggs, D. J. Sullivan, C. J. Brabec,
M. Buongiorno Nardelli, K. Rapcewicz, C. Roland, and M. Wensell,
Intl.\ J. Quant.\ Chem.\ {\bf 65}, 531 (1997). 

\bibitem{beck} 
M. P. Merrick, K. A. Iyer, and T. L. Beck, J. Phys.\ Chem.\
{\bf 99}, 12478 (1995);
K. A. Iyer, M. P. Merrick, and T. L. Beck, J. Chem.\ Phys.\
{\bf 103}, 227 (1995);
T. L. Beck, K. A. Iyer, M. P. Merrick, Intl.\ J. Quant.\ Chem.\
{\bf 61}, 341 (1997); 
T. L. Beck, {\it ibid.} 
{\bf 65}, 477 (1997). 

\bibitem{hamming}
R. W. Hamming, {\it Numerical Methods for Scientists and 
Engineers} (Dover, New York, 1962). Chapters 14 and 15. 

\bibitem{numrec} W.H. Press, B.P. Flannery, S.A. Teukolsky,
and W.T Vetterling, {\it Numerical Recipes in C: 
The Art of
Scientific Computing} (Cambridge Univ. Press, Cambridge,
1992).    

\end{thebibliography}
\end{document}